\documentclass[12pt,preprint]{aastex}

\usepackage{epsfig} 

\newcommand{\beq}{\begin{equation}}
\newcommand{\eeq}{\end{equation}}
\newcommand{\beqa}{\begin{eqnarray}}
\newcommand{\eeqa}{\end{eqnarray}}
\newcommand{\om}{\Omega_m}
\newcommand{\ok}{\Omega_k}
\newcommand{\obw}{\Omega_{\rm bw}}

\newcommand{\omde}{\Omega_{\rm DE}}
\newcommand{\ome}{\Omega_e}
\newcommand{\omv}{\Omega_\nu}
\newcommand{\oms}{\Omega_\star}
\newcommand{\omds}{\Omega_{\rm ds}}
\newcommand{\winf}{w_\infty}
\newcommand{\dls}{d_{\rm lss}}
\newcommand{\calr}{\mathcal{R}}
\newcommand{\lcdm}{\Lambda{\rm CDM}}
\newcommand{\lam}{\Lambda}

\newcommand{\ls}{\mathrel{\raise0.27ex\hbox{$<$}\kern-0.70em \lower0.71ex\hbox{{
$\scriptstyle \sim$}}}}

\begin{document} 

\title{Looking Beyond Lambda with the Union Supernova Compilation} 
\author{
D.~Rubin\altaffilmark{1,2},
E.~V.~Linder\altaffilmark{1,3},
M.~Kowalski\altaffilmark{4},
G.~Aldering\altaffilmark{1},
R.~Amanullah\altaffilmark{1,3},
K.~Barbary\altaffilmark{1,2},
N.~V.~Connolly\altaffilmark{5},
K.~S.~Dawson\altaffilmark{1},
L.~Faccioli\altaffilmark{1,3},
V.~Fadeyev\altaffilmark{6},
G.~Goldhaber\altaffilmark{1,2},
A.~Goobar\altaffilmark{7},
I.~Hook\altaffilmark{8},
C.~Lidman\altaffilmark{9},
J.~Meyers\altaffilmark{1,2},
S.~Nobili\altaffilmark{7},
P.~E.~Nugent\altaffilmark{1}, 
R.~Pain\altaffilmark{10},
S.~Perlmutter\altaffilmark{1,2},
P.~Ruiz-Lapuente\altaffilmark{11}, 
A.~L.~Spadafora\altaffilmark{1},
M.~Strovink\altaffilmark{1,2},
N.~Suzuki\altaffilmark{1}, and
H.~Swift\altaffilmark{1,2} \\ 
(Supernova Cosmology Project)
} 
\altaffiltext{1}{E. O. Lawrence Berkeley National Laboratory, 1 Cyclotron Rd., Berkeley, CA 94720, USA }
\altaffiltext{2}{Department of Physics, University of California Berkeley, Berkeley, 94720-7300 CA, USA}
\altaffiltext{3}{Space Sciences Laboratory, University of California Berkeley, Berkeley, CA 94720, USA}
\altaffiltext{4}{Humboldt Universit\"at Institut f\"ur Physik, Newtonstrasse 15, Berlin 12489, Germany}
\altaffiltext{5}{Department of Physics, Hamilton College, Clinton, NY 13323, USA}
\altaffiltext{6}{Department of Physics, University of California Santa Cruz, Santa Cruz, CA 95064, USA}
\altaffiltext{7}{Department of Physics, Stockholm University,  Albanova University Center, S-106 91 Stockholm, Sweden}
\altaffiltext{8}{Sub-Department of Astrophysics, University of Oxford, Denys Wilkinson Building, Keble Road, Oxford OX1 3RH, UK}
\altaffiltext{9}{European Southern Observatory, Alonso de Cordova 3107, Vitacura, Casilla 19001, Santiago 19, Chile }
\altaffiltext{10}{LPNHE, CNRS-IN2P3, University of Paris VI \& VII, Paris, France }
\altaffiltext{11}{Department of Astronomy, University of Barcelona, Barcelona, Spain} 


\begin{abstract}
The recent robust and homogeneous analysis of the world's supernova 
distance-redshift data, together with cosmic microwave background and 
baryon acoustic oscillation data, provides a powerful tool for constraining 
cosmological models.  Here we examine particular classes of scalar field, 
modified gravity, and phenomenological models to assess whether they 
are consistent with observations even when their behavior deviates from 
the cosmological constant $\Lambda$.  Some models have tension 
with the data, while others survive only by approaching the cosmological 
constant, and a couple are statistically favored over $\lcdm$.  Dark energy 
described by two equation of state 
parameters has considerable phase space to avoid $\Lambda$ and next 
generation data will be required to constrain such physics. 
\end{abstract} 



\section{Introduction \label{sec:intro}}

A decade after the discovery of the acceleration of the cosmic expansion 
\citep{perl99,riess98} we still understand little about the nature of 
the dark energy physics responsible.  Improved data continues to show 
consistency with Einstein's cosmological constant $\lam$, and in terms of a 
constant equation of state, or pressure to density, ratio $w$, the best 
fit to the data is $w=-0.969^{+0.059}_{-0.063}({\rm stat})^{+0.063}_{-0.066} 
({\rm sys})$, where $\Lambda$ has $w=-1$ \citep{union}.  
However, the magnitude of $\lam$ required and the coincidence 
for it becoming dominant so close to the present remain unexplained, and 
an abundance of motivated or unmotivated alternative models fills the 
literature.  Using the latest, most robust data available we examine 
the extent to which data really have settled on the cosmological constant. 

The vast array of models proposed for dark energy makes comparison of 
every model in the literature to the data a Sisyphean task.  Here we 
select some dozen models with properties such as well defined physical 
variables, simplicity, or features of particular physical interest.  
These embody a diversity of physics, including scalar fields, phase 
transitions, modified gravity, symmetries, and geometric relations. 
While far from exhaustive, they provide roadmarks for how well we can 
say that current data have zoomed in on $\lam$ as the solution. 

For such comparisons it is critical to employ robust data clearly 
interpretable within these ``beyond $\lam$'' cosmologies.  Geometric 
probes from the Type Ia supernovae (SN) distance-redshift relation, 
cosmic microwave background (CMB) acoustic peak scale shift parameter, 
and baryon acoustic oscillations (BAO) angular scale serve this essential 
role.  Equally important is confidence in the error estimates, 
incorporating systematics as well as statistical uncertainties.  
This has been studied in detail in the recent unified analysis of the world's 
published heterogeneous SN data sets -- the Union08 compilation \citep{union}. 

This SN compilation includes both the large data 
samples from the SNLS and ESSENCE survey, the compiled high redshift 
SNe observed with the Hubble Space Telescope, a new sample of nearby 
SNe, as well as several other, small data sets.  All SNe 
have been analyzed in a uniform manner and have passed a 
number of quality criteria (such as having data available in 
two bands to measure a color, 
and sufficient lightcurve points to make a meaningful fit).  
The samples have been carefully tested for inconsistencies under a 
blinded protocol 
before combining them into a single final data set comprising 307 SNe, the 
basis for this analysis. In this work the SNe data will be combined with 
the constraints obtained from the baryon acoustic oscillation scale 
\citep{eisenstein05} and from the five year data release of WMAP and 
ground based CMB measurements \citep{komatsu08}.  

In Section~\ref{sec:models} we describe the general method for 
cosmological parameter estimation and present a summary table of 
the various models considered and the $\chi^2$ statistics of the fit.  
Sections~\ref{sec:const}--\ref{sec:grownu} then briefly describe 
the dark energy models, their parameters, and show the likelihood 
contours.  The concluding discussion occurs in Section~\ref{sec:concl}.

\section{Constraining Models \label{sec:models}} 

Achieving informative constraints on the nature of 
dark energy requires restricting the degrees of freedom of the theory 
and the resulting degeneracies in the cosmological 
model being tested.  One degree of freedom entering the model is the 
present matter density $\om$.  For the case of the spatially flat 
cosmological constant $\Lambda$ model (or some of the other models 
considered below), this is the sole cosmological parameter determining the 
distances entering the supernova (SN) magnitude-redshift, baryon acoustic 
oscillation scale (BAO), and cosmic microwave background (CMB) shift 
parameter relations. 

Generally, further degrees of freedom to describe the nature of the dark 
energy, i.e.\ its equation of state (EOS), or pressure to density, ratio, 
are needed.  In a few cases the EOS is parameter free, as in the $\Lambda$ 
case where $w=-1$, or is determined by the matter density, as in some  
subcases below (such as the flat DGP braneworld gravity model of 
\S\ref{sec:dgp}).  One way to categorize models is by the number of 
independent EOS parameters, or 
general parameters beyond the matter density (so flat $\Lambda$ models 
have zero such parameters, $\Lambda$ models with curvature have one).  
In general, current data can deliver reasonable constraints on one 
parameter descriptions of dark energy. 

In addition to exploring the nature of dark energy through its EOS, one 
might also include another parameter for the dark energy density, 
i.e.\ allow the possibility of nonzero spatial curvature.  In this case 
individual probes then generally do a poor job constraining the model 
with current data, although the combined data from SN+CMB+BAO can sometimes 
still have leverage.  Since crosschecks and testing consistency between 
probes is important (as particularly illustrated below in the DGP case), 
we consider spatial curvature only in the otherwise zero parameter cases 
of $\Lambda$ and DGP, and for the constant EOS dark energy model. 

In the following sections we investigate various one parameter EOS models, 
discussing their physical motivation or lack thereof, and features of 
interest, and the observational constraints that can be placed upon them.  
In the last sections we also investigate some two parameter models of 
interest, with constrained physical behaviors and particular motivations.  
As a preview and summary of results, Table~\ref{tab:chi2} 
lists the models, number of parameters, and goodness of fit 
for the present data.  

The SN, CMB, and BAO data are combined by multiplying the likelihoods. 
Especially when testing models deviating from the cosmological constant 
one must be careful to account for any shift of the CMB sound horizon 
arising from violation of high redshift matter domination on the CMB 
and BAO scales; details are given in Union08.  Note that some doubt 
exists on the use of the BAO constraints for cosmologies other than 
$\lcdm$, or possibly constant $w$, \citep{dickknoxchu,goobar07} since  
$\lcdm$ is assumed in several places in the \citet{eisenstein05} analysis, 
e.g.\ computation of the correlation function from redshift space, 
nonlinear density corrections, structure formation and the matter power 
spectrum, and color and luminosity function evolution.  Properly, a 
systematic uncertainty should be assigned to BAO to account for these 
effects; however, this requires a complex analysis from the original data 
and we show only the statistical error.  At the current level of precision, 
simplified estimates show this does not strongly affect the results, but 
such systematics will need to be treated for future BAO data. 
All figures use 
the likelihood maximized over all relevant parameters besides those 
plotted, and contours are at the 68.3\%, 95.4\%, and 99.7\% confidence 
level.

\begin{table}[htbp]
\begin{center}
\begin{tabular}{ l l l r@{.}l r@{.}l  }
\hline
\hline
Model & Motivation & Parameters & \multicolumn{2}{l}{$\chi^2$ (stat)} & 
\multicolumn{2}{l}{$\chi^2$ (sys)}\\
\hline
$\Lambda$CDM  (flat) &gravity, zeropoint&$\om$ & 313&1 & 309&9 \\
\hline
& & &  \multicolumn{2}{l}{$\Delta \chi^2$ (stat)} & 
\multicolumn{2}{l}{$\Delta\chi^2$ (sys)}\\
$\Lambda$CDM &gravity, zeropoint& $\om$, $\Omega_\Lambda$ & $-$1&1 & $-1$&3 \\
Constant  $w$ (flat)&simple extension& $\om$, $w$ &$-0$&3 & $-1$&2 \\
Constant  $w$  &simple extension& $\om$, $\Omega_k, w$ &$-$1&1 & $-1$&6 \\

\hline

Braneworld &consistent gravity& $\om$, $\Omega_k$ & 15&0 & 2&7 \\
Doomsday &simple extension& $\om$, $t_{\rm doom}$ & $-$0&1 & $-0$&7 \\
Mirage &CMB distance& $\om$, $w_0$ & $-$0&2 & $-0$&1\\
Vacuum Metamorphosis &induced gravity& $\om$, $\Omega_\star$ & 0&0 & 0&0 \\
Geometric DE ${\rm R_{low}}$ &kinematics& $r_0,r_1$ $(\om,w_0)$ & 0&1& $-1$&1\\
Geometric DE ${\rm R_{high}}$ &matter era deviation& $\om$, $w_\infty$, $\beta$ & $-$1&9 &$-2$&2\\
PNGB &naturalness& $\om$, $w_0$, $f$ &$-0$&1 & $-0$&7 \\
Algebraic Thawing &generic evolution& $\om$, $w_0$, $p$ & $-1$&6 & $-2$&3 \\
Early DE &fine tuning problem& $\om$, $w_0$, $\Omega_e$ & $-0$&3 & $-1$&2 \\
Growing $\nu$-mass &coincidence problem& $\om$, $\Omega_e$, $m_\nu^0$& $-0$&6 & $-1$&6\\
\hline
\end{tabular}
\caption{``Beyond $\Lambda$'' dark energy models considered in this paper,
together with $\Lambda$CDM models.  Models are listed in the order of
discussion, and the cosmological fitting parameters shown.  The $\chi^2$
of the matter plus cosmological constant case is given, and all other models
list the $\Delta\chi^2$ from that model.  The values refer to the best fit
to the joint data of SN+CMB+BAO; in the last column the SN systematics as
analyzed in Union08 are included.  }
\label{tab:chi2}
\end{center}
\end{table}

It is particularly important to note the treatment of systematic 
errors, included only for SN.  We employ the prescription of Union08 
for propagation of systematic errors.  This introduces a new distance 
modulus $\mu^{\rm sys} = \mu+\Delta M_i+\Delta M$, which is simply the 
usual distance modulus $\mu=5\log(H_0 d_L(z))$, 
where $d_L(z)$ is the luminosity distance and 
$H_0$ the Hubble constant, shifted by a sample dependent
magnitude offset $\Delta M_i$ and a 
single sample independent magnitude offset $\Delta M$ added only for 
the higher redshift SNe ($z>0.2$).  
The magnitude offsets $\Delta M_i$ reflect possible 
heterogeneity among the SNe samples while the $\Delta M$ step from 
SNe at $z<0.2$ to $z>0.2$ allows a possible common systematic error 
in the comparison of low vs.\ high redshift SNe. 
Treating $\Delta M_i$ and $\Delta M$ as additional 
fit parameters, one defines $\chi_{\rm sys}^2=\chi^2+\sum_i 
(\Delta M_i/\sigma_{M_i})^2 + (\Delta M/\sigma_{M})^2$ to absorb the 
uncertainty in the nuisance 
parameters, $\sigma_{M_i}$ and $\sigma_{M}$, and obtain constraints 
on the desired physical fit parameters that include systematic errors. 
This procedure of incorporating systematic errors provides robust 
quantification of whether or not a model is in conflict with the data 
and is essential for accurate physical interpretation. 
See Union08 for further, detailed discussion of robust treatment of 
systematics within the current world heterogeneous SN data.

\section{Constant Equation of State \label{sec:const}} 

Models with constant equation of state $w$ within 20\%, say, of the 
cosmological constant value $w=-1$, but not equal to $-1$, do not 
have much physical motivation.  To achieve a constant equation of state 
requires fine tuning of the kinetic and potential energies of a scalar 
field throughout its evolution.  It is not clear that a constant $w\ne-1$ 
is a good approximation to any reasonable dynamical scalar field, where 
$w$ varies, and certainly does not capture the key physics.  However, 
since current data cannot discern EOS variation on timescales less than 
or of order the Hubble time, traditionally one phrases constraints in terms 
of a constant $w$.  We reproduce this model from Union08 to serve as 
a point of comparison.  Also see Union08 for models using the 
standard time varying EOS $w(a)=w_0+w_a(1-a)$, where $a=1/(1+z)$ is the 
scale factor, and models with $w(z)$ given in redshift bins. 

In the constant $w$ case the Hubble expansion parameter 
$H=\dot a/a$ is given by 
\beq 
H^2(z)/H_0^2 =\Omega_m(1+z)^3+\Omega_w(1+z)^{3(1+w)}+\Omega_k(1+z)^2, 
\eeq 
where $\om$ is the present matter density, $\Omega_w$ the present 
dark energy density, and $\Omega_k=1-\om-\Omega_w$ the effective energy 
density for spatial curvature.  

\begin{figure}[!ht]
\begin{center}
\psfig{file=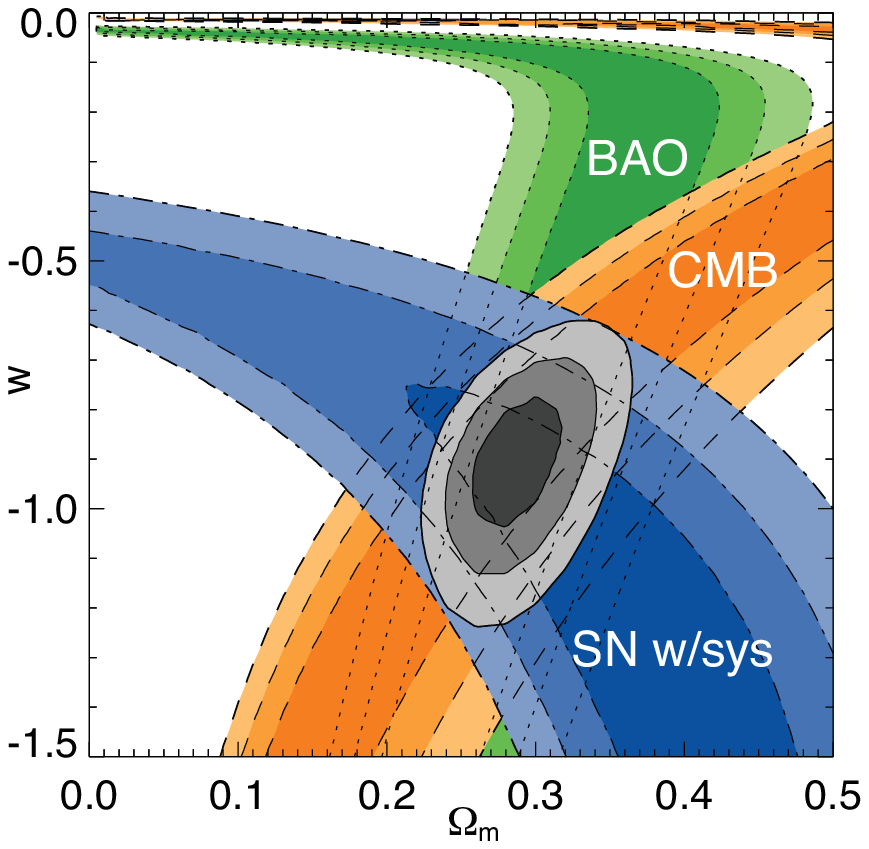, width=3.2in} 
\psfig{file=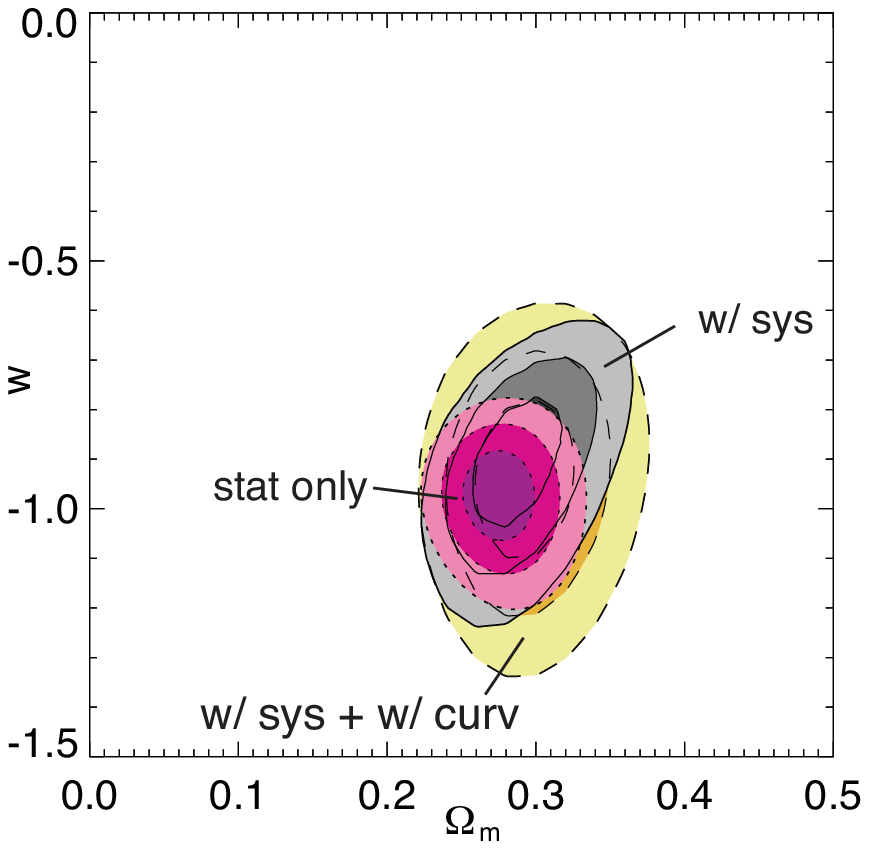, width=3.2in} 
\caption{68.3\%, 95.4\%, and 99.7\% confidence level contours on a constant 
EOS $w$ and the matter density $\om$ for the individual and combined data 
sets.  The left panel shows individual and combined probes in the flat 
universe case; the right panel repeats the combined systematics contour, 
and also compares to the statistical only contour, and to the systematics 
contour when simultaneously fitting for spatial curvature.  
} 
\label{fig:wconst}
\end{center} 
\end{figure}

Figure~\ref{fig:wconst} shows the confidence contours in the $w$-$\om$ 
plane both without and with (minimized in the likelihood fit) spatial 
curvature.  Note that allowing for spatial curvature does not 
strongly degrade the constraints. This is due to the strong 
complementarity of SN, CMB, and BAO data, combined with the restriction 
to a constant $w$ model.  As shown in Union08, the constraint 
on curvature in this model is $\Omega_k=-0.010\pm0.012$.  
See Union08 for more plots showing the individual probe constraints.

\section{Braneworld Gravity \label{sec:dgp}} 

Rather than from a new physical energy density, cosmic acceleration could 
be due to a modification of the Friedmann expansion equations arising 
from an extension of gravitational theory.  
In braneworld cosmology \citep{dgp,ddg}, the acceleration is caused by 
a weakening of gravity over distances near the Hubble scale due to leaking 
into an extra dimensional bulk from our four dimensional brane.  Thus a 
physical dark energy is replaced by an infrared modification of gravity.  
For DGP braneworld gravity, the Hubble expansion is given by 
\beqa 
H^2/H_0^2&=&\left(\sqrt{\om(1+z)^3+\obw}+\sqrt{\obw}\right)^2+ 
\Omega_k(1+z)^2 \\ 
&\rightarrow&\om(1+z)^3+2\obw+2\sqrt{\obw}\sqrt{\om(1+z)^3+\obw}, 
\qquad {\rm (flat)}\,.  
\eeqa 
Here the present effective braneworld energy density is 
\beqa 
\obw&=&\frac{(1-\om-\ok)^2}{4(1-\ok)} \\ 
&\rightarrow&\frac{(1-\om)^2}{4}, \qquad {\rm (flat)}\,, 
\eeqa 
and is related to the five dimensional crossover scale 
$r_c=M_{\rm Pl}^2/(2M_5^3)$ by $\obw=1/(4H_0^2r_c^2)$.  
Note that the 
only cosmological parameters for this model are $\om$ and $\ok$ (or 
$\obw$), so it has the same number of parameters as $\lcdm$. 

\begin{figure}[!ht]
\begin{center}
\psfig{file=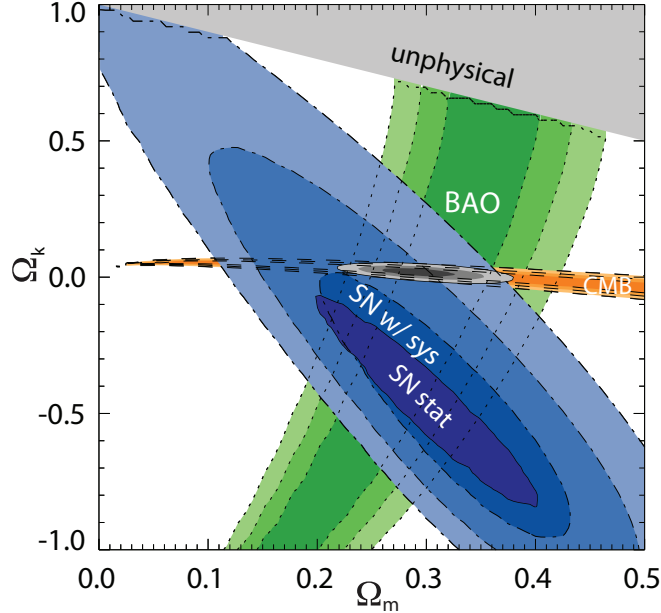, width=3.4in} 
\caption{The extradimensional DGP braneworld gravity model does not 
achieve an acceptable fit to the combined data, even allowing for a 
spatial curvature parameter.  The joint best fit is in fact a nearly flat 
model, but with poor goodness of fit: $\Delta\chi^2=2.7$ relative to the 
$\lcdm$ case; also shown is the statistical error only SN contour, 
which gives a joint $\Delta\chi^2=15$ relative to $\lcdm$. 
} 
\label{fig:dgp}
\end{center} 
\end{figure}

The effective dark energy equation of state is given by the simple 
expression 
\beq 
w(z)=-\frac{1-\ok(z)}{1+\om(z)-\ok(z)}, \label{eq:bww}
\eeq 
where $\om(z)=\om(1+z)^3/(H^2/H_0^2)$ and $\ok(z)=\ok(1+z)^2/(H^2/H_0^2)$.  
Thus the dark energy equation of state at present, $w_0$, is determined 
by $\om$ 
and $\ok$; while time varying, it is not an independent parameter.  So 
rather than plotting $w_0$ vs.\ $\om$ or showing constraints on the somewhat 
nonintuitive parameters $r_c$ or $\obw$ (but see the clear discussion and 
plots in \citet{davis07,goobar07}, though without systematics), 
Figure~\ref{fig:dgp} illustrates 
the confidence contours in the $\ok$-$\om$ plane.  This makes it 
particularly easy to see how deviations from flatness pull the value 
of the matter density. 
In this and following figures, dotted contours show the BAO constraints, 
dashed for CMB  constraints, dot-dashed for SN with systematics, and 
solid contours give the joint constraints.

For a flat universe, in order for $w$ to approach $-1$ the matter density 
is forced to small values.  Alternately, pushing the curvature density 
$\ok$ negative, i.e.\ introducing a positive spatial curvature $k$, 
allows $w\approx-1$ with higher matter density.  For a given $w_0$, the 
amount of curvature needed 
can be derived from Eq.~(\ref{eq:bww}) to be approximately $\Delta\ok 
\approx -\Delta\om/\om$, so to move a flat, $\om=0.2$ universe to $\om=0.3$ 
requires $\ok=-0.5$, in agreement with the SN contour (being most sensitive 
to $w_0$) of Figure~\ref{fig:dgp}. 

Note that the curvature density cannot exceed $1-\om$, corresponding to 
an infinite crossover scale $r_c$, so the likelihood contours are cut 
off at this line and the region beyond is unphysical.  However, this 
does not affect the joint contours. 
The BAO data contours do extend to the limit $\ok=1-\om$; here 
$\obw=0$, equivalent to the simple OCDM open, nonaccelerating universe. 

Most importantly, the three probes do not reach concordance on a given 
cosmological model.  The areas of intersection of any pair are distinct 
from other pairs, indicating that the full data disfavors the braneworld 
model, even with curvature.  This is further quantified by the poor 
goodness of fit to the data, with $\Delta\chi^2=2.7$ relative to the 
flat $\lcdm$ model possessing one fewer parameter, or $\Delta\chi^2=4.0$ 
relative to $\lcdm$ allowing curvature. This indicates the crucial 
importance of crosschecking probes.  Moreover, if we had used only the 
statistical estimates of uncertainties (see the ``SN stat'' 68\% cl 
contour of Fig.~\ref{fig:dgp}), 
we would have found that $\Delta\chi^2=15$ rather than 2.7, and possibly 
drawn exaggerated physical conclusions -- considering the DGP model 2000 
times less likely than it really is, as an illustration\footnote{This 
is not 
quite fair as the braneworld model and $\lcdm$ model have distinct 
parameter spaces and the reduced $\chi^2/$ dof is only 1.07 for the 
statistics only braneworld case.  This is one area where Bayesian 
evidence methods, with careful use of priors, would be useful for model 
comparison.}.  
Inclusion of 
systematics is essential for robust interpretation of results.

\section{Doomsday Model} 

Perhaps the simplest generalization of the cosmological constant is the 
linear potential model, pioneered by \citet{linde86} and discussed 
recently by \citet{wbgbook}, motivated from high energy physics.  
Interestingly, while this gives a current accelerating epoch, in the 
future the potential becomes negative and not only deceleration of the 
expansion but collapse of the universe ensues.  Hence the name of doomsday 
model.  

The potential has two parameters: the 
amplitude and slope.  The amplitude $V_0$ essentially gives the 
dark energy density, which is fixed by $\om$ in a flat universe.  
(For the remainder of the paper we assume a flat universe, for the 
reasons discussed in \S\ref{sec:models}.)  
The slope $V'=dV/d\phi$ can be translated into the present equation of 
state value $w_0$.  Thus this is a one parameter model in our categorization.  
See \citet{0307185} for discussion of the cosmological properties 
of the linear potential, \citet{linde86} for a view of it as a perturbation 
about zero cosmological constant, and \citet{0307004} for links to the 
large kinetic 
term approach in particle physics.  More recently, this has been 
considered as a textbook case by \citet{wbgbook}, so we will examine 
this model in some detail.  Such dark energy is an example of a thawing 
scalar field \citep{caldlin}, starting 
with $w(z\gg1)=-1$ and slowly rolling to attain less negative values of 
$w$; that is, it departs from $\lam$.  If it has not evolved too far 
from $-1$ then its behavior is well described by $w_a\approx-1.5(1+w_0)$ 
where $w(a)=w_0+w_a(1-a)$. 
However we solve the scalar field equation of motion exactly for 
all results quoted here. 

As the scalar field rolls to small values of the potential 
the expansion stops accelerating, and when it reaches $V=0$ then 
$w=1$.  However it crosses through zero to negative values of the potential, 
further increasing $w$, and eventually the dark energy density itself 
crosses through zero, causing $w$ to go to positive and then negative 
infinity.  Thereafter the negative dark energy 
density, acting now as an attractive gravitational force, causes not 
only deceleration but forces the universe to start 
contracting.  The rapid collapse of the universe ends in a Big Crunch, or 
cosmic doomsday in a finite time.  

\begin{figure}[!ht]
\begin{center}
\psfig{file=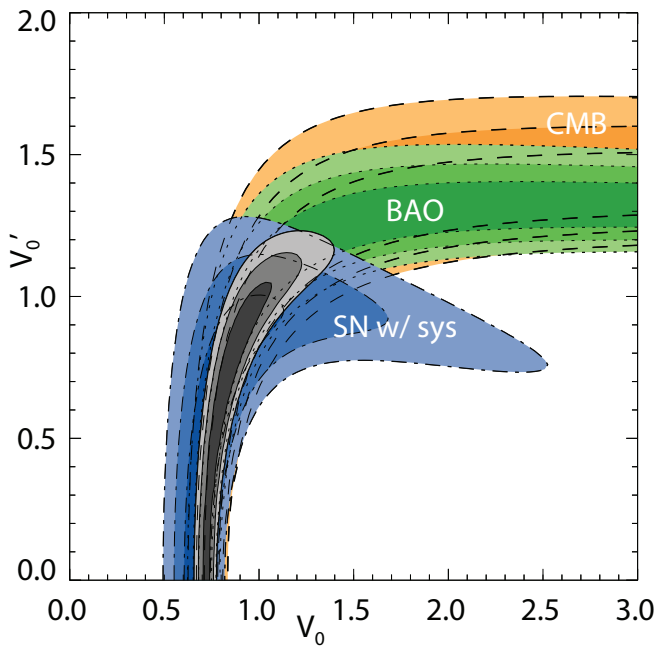, width=3.4in} 
\caption{Constraints on the linear potential model in terms of the 
high energy physics quantities of the primordial amplitude and slope 
of the potential.  Note there is less complementarity between some 
of the probes than for the constant $w$ model.  Fig.~\ref{fig:doom} 
translates these constraints into ones on the cosmological parameters. 
} 
\label{fig:doomhep}
\end{center} 
\end{figure}

In the notation used in \citet{wbgbook}, $V(\phi)=V_0+(\phi-\phi_0)\,V'_0$, 
with $V_0$ the potential energy during the initial frozen state (during 
high Hubble drag at high redshift) and $V'_0$ is the 
constant potential slope.  Figure~\ref{fig:doomhep} shows the constraints 
in this high energy physics plane $V_0$-$V'_0$.  Note the tight constraints 
on the initial potential energy $V_0$, given in units of the present 
critical density.  The cosmological constant corresponds to the limit of 
$V'_0=0$, but the slope must always be less than or of order $10^{-120}$ 
in Planck units, i.e.\ unity when shown in terms of the present energy 
density, to match the data.

\begin{figure}[!ht]
\begin{center}
\psfig{file=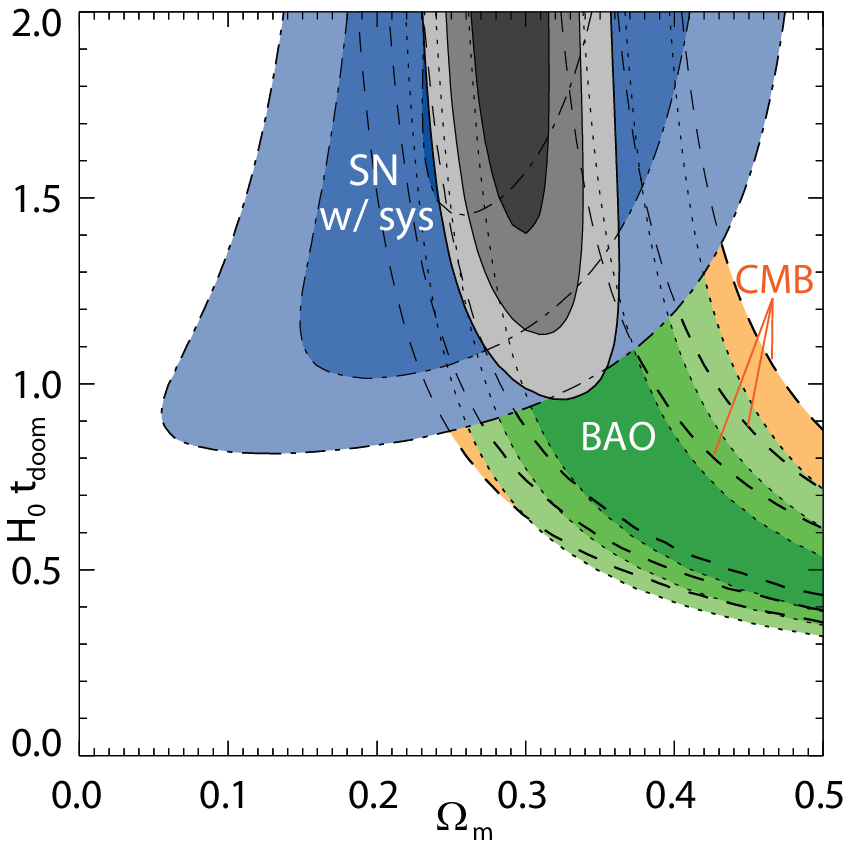, width=3.2in} 
\psfig{file=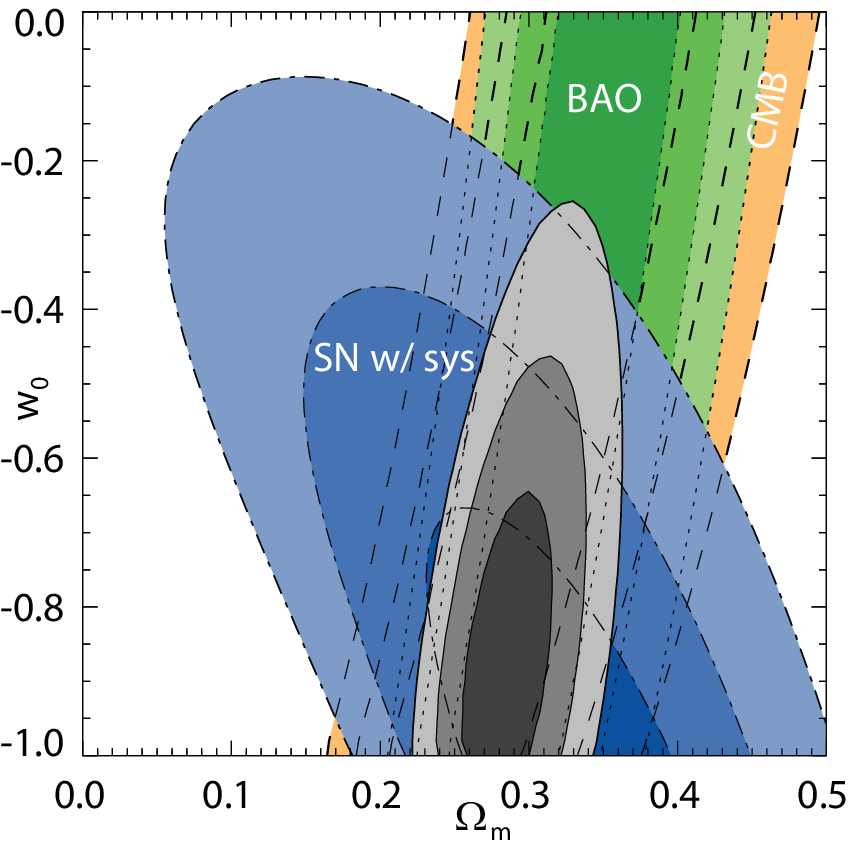, width=3.2in} 
\caption{The future expansion history in the linear potential model 
has a collapse, or cosmic doomsday, at a finite time in the future.  
The left panel shows the confidence 
contours for the time remaining until collapse; the likelihood contours 
extend to infinity, with $t_{\rm doom}=\infty$ corresponding to the 
$\lam$ model.  The contours can also be viewed in the equivalent 
$w_0$-$\om$ plane (right panel).  
Current data constraints indicate cosmic doomsday will occur no sooner 
than $\sim$1.24 Hubble times from now at 95\% confidence.  
} 
\label{fig:doom}
\end{center} 
\end{figure}

We can also translate these high energy physics parameters 
into the recent universe quantities of the matter density $\om$ and the 
present equation of state $w_0$.  Moreover, this is directly related 
to the doomsday time $t_{\rm doom}$, or future time until collapse. 
A useful approximation (though we employ the exact solution) between 
$t_{\rm doom}$, $w_0$, and the approximate time variation $w_a=-1.5(1+w_0)$ 
is 
\beq 
t_{\rm doom}\approx 0.5 H_0^{-1}(1+w_0)^{-0.8}\approx 
0.6H_0^{-1}(-w_a)^{-0.8}. 
\eeq 
Figure~\ref{fig:doom} shows the likelihood contours in the 
$t_{\rm doom}$-$\om$ and $w_0$-$\om$ planes.  
The 95\% confidence limit on $t_{\rm doom}$ from present observations is 
$1.24\,H_0^{-1}$, i.e.\ we are 95\% likely to have at least 
17 billion more years before doomsday!

\section{Mirage Model} 

Given their limited sensitivity to the dynamics of dark energy, current data 
can appear to see a cosmological constant even in the presence of time 
variation.  This is called the ``mirage of $\Lambda$'', and we consider 
mirage models, with a form motivated by the observations as discussed below, 
specifically to test whether the concordance cosmology truly narrows in 
on the cosmological constant as the dark energy.  

Since cosmological distances involve an integral over the energy density 
of components, which in turn are integrals over the equation of state 
as a function of redshift, there exists a chain of dependences between 
these quantities.  Fixing a distance, such as $\dls$ to the CMB last 
scattering surface, can generally lead to an ``attractor'' behavior in 
the equation of state to a common averaged value or the value at a 
particular redshift. 
Specifically, \citet{wcon} pointed out that if CMB data 
for $\dls$ is well fit by the $\lcdm$ model then this forces 
$w(z\approx0.4)\approx-1$ for quite general monotonic EOS.  
So even dark energy models with substantial 
time variation could thus appear to behave like the cosmological constant 
at $z\approx0.4$, near the pivot redshift of current data.  

Since current 
experiments insensitive to time variation inherently interpret the data 
in terms of a constant $w$ given by the EOS value at the pivot redshift, 
this in turn thus leads to the ``mirage of $\Lambda$'': thinking that 
$w=-1$ everywhere, despite models very different from $\Lambda$ being 
good fits.  See \S5.2 of \citet{LinRPP} for further discussion. 
(Also note that attempting to constrain the EOS by combining the CMB 
$\dls$ with a precision determination of the Hubble constant $H_0$ 
only tightens the uncertainty on the pivot equation of state value 
(already taken to be nearly $-1$) and so similarly does not reveal 
the true nature of dark energy.) 

We test this with a family of ``mirage'' models motivated by the 
reduced distance to CMB last scattering $\dls$.  These correspond to 
the one parameter subset of the two parameter EOS model 
$w(a)=w_0+w_a(1-a)$ with $w_a$ determined by 
$w_a=-3.63(1+w_0)$.  They are not exactly equivalent to imposing a 
CMB prior since $d_{\rm lss}$ will still change with $\om$; that is, 
they essentially test the uniqueness of the current concordance model 
for cosmology: $\lcdm$ with $\om=0.28$.  

For any model well approximated by a relation $w_a=-A(1+w_0)$, as 
this model (and the previous one) is, the Hubble parameter is given by 
\beqa 
H^2/H_0^2&=&\om\,(1+z)^3+(1-\om)\,(1+z)^{3(1+w_0+w_a)}\,e^{-3w_a z/(1+z)} \\ 
&=&\om\,(1+z)^3+(1-\om)\,(1+z)^{3(1+w_0)(1-A)}\,e^{3A(1+w_0) z/(1+z)} 
\,. \label{eq:huba}
\eeqa 

\begin{figure}[!ht]
\begin{center}
\psfig{file=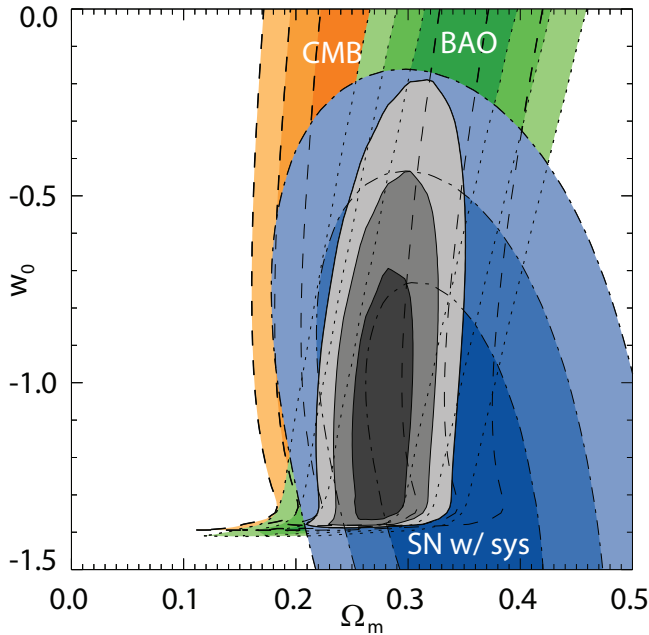, width=3.4in} 
\caption{The mirage subclass of time varying dark energy looks like 
$\Lambda$ in an averaged sense.  Note that CMB contours are 
almost vertical, indicating both that the mirage holds, preserving the 
$\lcdm$ distance to last scattering, and yet imposes little constraint on 
$w_0$, and hence $w_a$.  Thus the appearance of $\Lambda$ does not 
actually exclude time variation.  The mirage is broken 
when the equation of state at high redshift exceeds the matter 
domination value of zero; this causes 
the wall in the likelihood at $w_0=A/(1-A)\approx-1.4$. 
} 
\label{fig:mirage}
\end{center} 
\end{figure}

Figure~\ref{fig:mirage} shows constraints in the $w_0$-$\om$ plane. 
It is important to note that $w$ is not constant in this model.  A 
significant range of $w_0$ (and hence a larger range of $w_a$ too, 
roughly $+0.55$ to $-1.1$ at 68\% cl) is 
allowed by the data, even though these models all look in an averaged 
sense like a cosmological constant.  Thus experiments sensitive to 
the time variation $w_a$ (e.g.\ $\sigma(w_a)<0.36$ to know that $w(z)$ 
is really, not just apparently, within 10\% of $-1$) are required to 
determine whether the mirage is reality or not. 

\section{Vacuum Metamorphosis \label{sec:vm}} 

An interesting model where the cosmic acceleration is due to a change in 
the behavior of physical laws, rather than a new physical energy density, 
is the vacuum metamorphosis model \citep{parkerraval,caldwellkpz}.  
As in Sakharov's induced gravity \citep{sakharov}, quantum fluctuations 
of a massive scalar field give rise to a phase transition in gravity when 
the Ricci scalar curvature $R$ becomes of order the mass squared of the 
field, and freezes $R$ there.  This model is interesting in terms 
of its physical origin and nearly first principles derivation, and further 
because it is an example of a well behaved phantom field, with $w<-1$. 

The criticality condition 
\beq 
R=6(\dot H+2H^2)=m^2 
\eeq 
after the phase transition at redshift $z_t$ leads to a Hubble parameter 
\beqa 
H^2/H_0^2&=& 
(1-\frac{m^2}{12})(1+z)^4+\frac{m^2}{12}\,, \qquad z<z_t\,, \\ 
H^2/H_0^2&=&\om(1+z)^3+\frac{m^2}{3}\frac{1-\oms}{4-3\oms}\,, \qquad z>z_t\,. 
\eeqa

There is one parameter, $\oms=\Omega_m(z_t)$, in addition to the present 
matter density $\om$, where $1-\oms$ is proportional to the cosmological 
constant.  The variables $z_t$ and $m$ are given in terms of 
$\om$, $\oms$ by 
$z_t=(m^2\oms/[3\om(4-3\oms)])^{1/3}-1$ and 
$m^2=3\om[(4-3\oms)/\oms]^{1/4}[(4/m^2)-(1/3)]^{-3/4}$.  
The original version of the model had fixed $\oms=1$, 
i.e.\ no cosmological constant, but if the scalar field has nonzero 
expectation value (which is not required for the induced gravity phase 
transition) then there will be a cosmological constant, and $\oms$ 
deviates from unity.  

\begin{figure}[!ht]
\begin{center}
\psfig{file=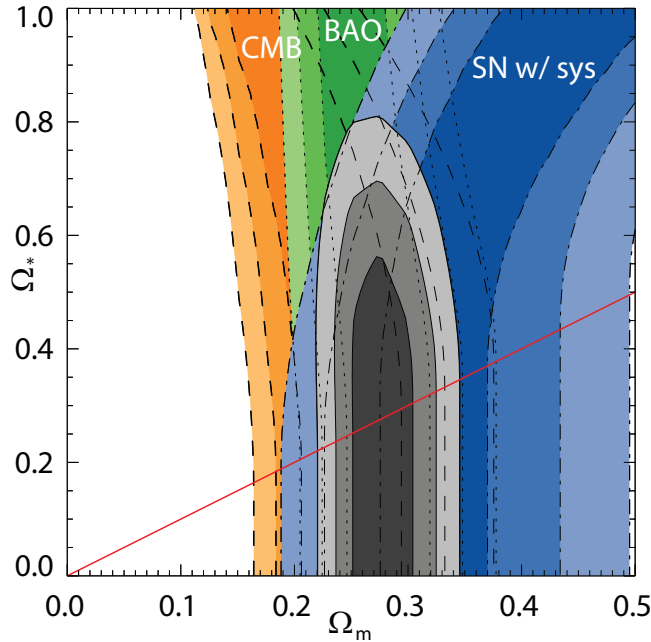, width=3.4in} 
\caption{The vacuum metamorphosis model involves a phase transition in 
gravitational laws due to quantum effects.   Where the quantum field 
inducing the gravitational deviation has no additional zeropoint energy, 
i.e.\ cosmological constant, then $\oms=1$, and the data gives discordant 
results.  As the model approaches 
the $\oms=\om$ line of pure cosmological constant plus matter without a phase 
transition in the past, the data provide an increasingly good fit.  (Below 
the line, the transition takes place further into the future, with no 
effect on the data likelihood.) 
} 
\label{fig:vacuum}
\end{center} 
\end{figure}

Figure~\ref{fig:vacuum} shows the confidence contours in the $\oms$-$\om$ 
plane.  To consider constraints on the original vacuum metamorphosis model, 
without an extra cosmological constant, slice across the likelihood contours 
at the $\oms=1$ line.  We see that the three probes are inconsistent with 
each other in this case, with disjoint contours (indeed the 
$\Delta \chi^2=28.5$ relative to flat $\lcdm$).  Allowing for a 
cosmological constant, i.e.\ $\oms\ne1$, brings the probes into 
concordance, and the best joint fit approaches the lower bound of the 
region $\oms\ge\om$.  The condition $\oms=\om$ corresponds 
to the standard cosmological constant case, with $\Omega_\lam=1-\om$, 
since the phase transition then only occurs at $z_t=0$.  Thus the data do 
not favor any vacuum phase transition. 
Although this model comprises very different physics, and allows phantom 
behavior, the data still are consistent with the cosmological constant.

\section{Geometric Dark Energy} 

According to the Equivalence Principle, acceleration is manifest 
in the curvature of spacetime, so it is interesting to consider geometric 
dark energy, the idea that the acceleration arises from some property 
of the spacetime geometry.  One example of this involves 
the holographic principle of quantum field theory as applied to cosmology. 
This limits the number of modes available to the vacuum energy and so could 
have an impact on the cosmological constant problem \citep{bousso}.  
The basic idea is that there is a spacelike, two 
dimensional surface on which all the field information is holographically 
encoded, and the covariant entropy bound relates the area of this surface 
to the maximum mode energy allowed (UV cutoff).  The vacuum 
energy density resulting from summing over modes ends up being 
proportional to the area, or inverse square of the characteristic 
length scale.   However, what is perhaps the natural surface to choose, 
the causal event horizon, does not lead to an energy density with 
accelerating properties.  

Many of the attempts in the literature to 
overcome this have grown increasingly distant from the original concept 
of holography, though they often retain the name.  It is important to 
realize that, dimensionally, any energy density, including the vacuum 
energy density, has $\rho\sim L^{-2}$, so merely choosing some length 
$L$ does not imply any connection to quantum holography.  We therefore 
do not consider these models but turn instead to the spacetime curvature.

\subsection{Ricci dark energy ${\rm R_{low}}$} 

A different approach involves the spacetime curvature directly, as measured 
through the Ricci scalar.  This is similar in motivation to the vacuum 
metamorphosis model of \S\ref{sec:vm}.  Here we consider it purely 
geometrically, with the key physical 
quantity being the reduced scalar spacetime curvature, in terms of the 
Ricci scalar and Hubble parameter, as in the model of \citet{LinGeom}, 

\beq 
\calr\equiv \frac{R}{12H^2} =r_0+r_1(1-a). \label{eq:ricci} 
\eeq 
This quantity directly involves the acceleration.  Moreover, we can 
treat it purely kinematically, as in the last equality above, assuming no 
field equations or dynamics.  Of course, any functional form contains 
an implicit dynamics (see, e.g., \citet{LinRPP}), but we have chosen 
effectively a Taylor expansion in the scale factor $a$, valid for any 
dynamics for small deviations $1-a$ from the present, i.e.\ the low 
redshift or low scalar curvature regime. 

At high redshift, as $1-a$ is no longer small, we match it onto an 
asymptotic matter dominated behavior for $a<a_t=1-(1-4r_0)/(4r_1)$.  
Solving for the Hubble parameter, 
\beqa 
H^2/H_0^2&=&a^{4(r_0+r_1-1)}\,e^{4r_1(1-a)}, \quad a>a_t \\ 
H^2/H_0^2&=&\om\,a^{-3}, \quad a<a_t\,. 
\eeqa 
The matching condition determines 
\beq 
\om=\left(\frac{4r_0+4r_1-1}{4r_1}\right)^{4r_0+4r_1-1}\,e^{1-4r_0}, 
\eeq 
so there is only one parameter independent of the matter density. 

\begin{figure}[!ht]
\begin{center}
\psfig{file=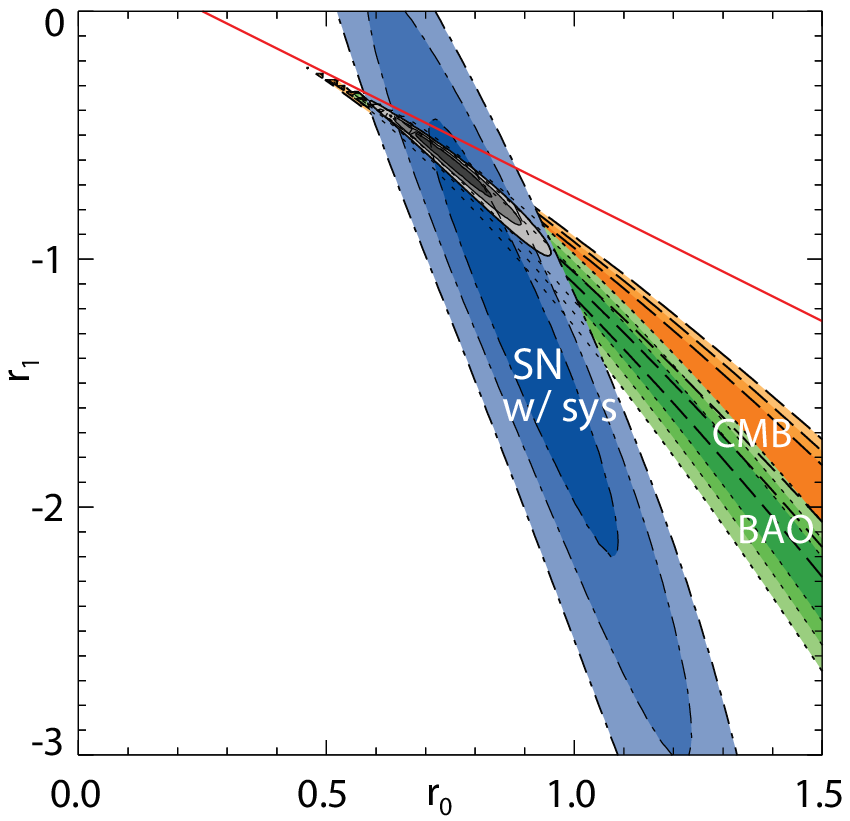, width=3.2in}
\psfig{file=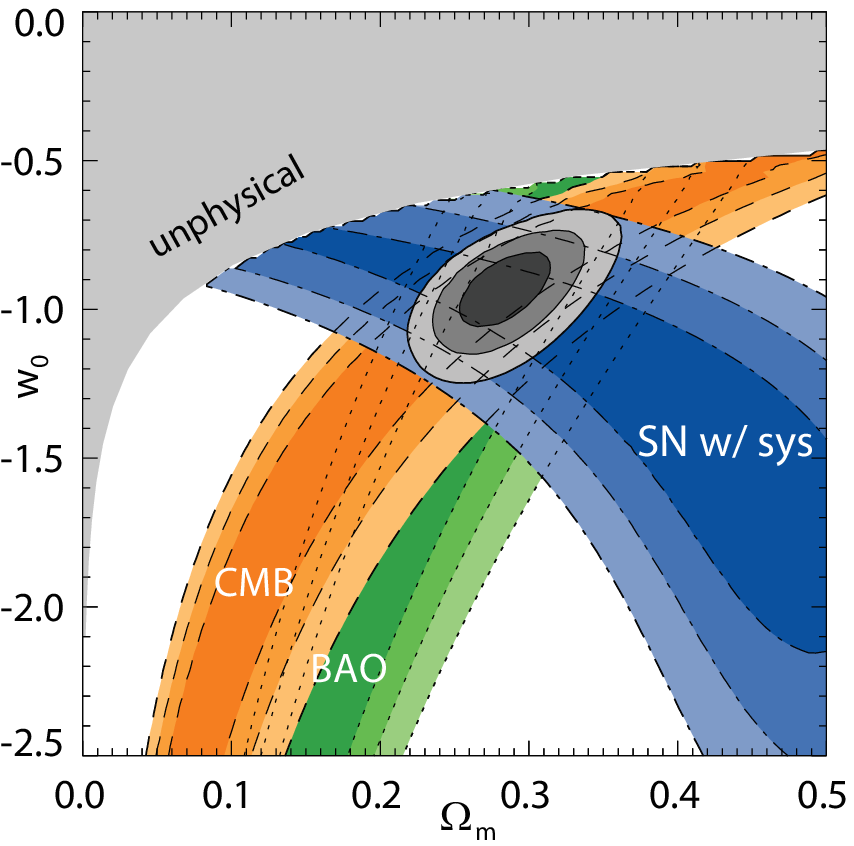, width=3.2in}
\caption{Geometric dark energy in the ${\rm R_{low}}$ model describes 
the acceleration directly through the reduced Ricci scalar, or spacetime 
curvature.  This can be viewed in a kinematic sense, in the $r_0$-$r_1$ 
plane, or in a dark energy sense in the $\om$-$w_0$ plane.  The data 
favor $w_0=-1$ but this is not $\Lambda$, instead representing distinct 
physics.  
For $r_0+r_1>1/4$, above the diagonal line, early matter domination is 
violated, and the CMB and BAO likelihoods avoid this region, as seen in 
the left panel; the matter density also cannot then be uniquely defined 
so the equivalent region is excluded from the right panel. 
}
\label{fig:geom1}
\end{center}
\end{figure}

Note also that we can define an effective dark energy as that part of 
the Hubble parameter deviating from the usual matter behavior, with 
equation of state generally given by 
\beq 
w(a)=\frac{1-4\calr}{3}\,\left[1-\om\,e^{-\int_a^1 (da/a)(1-4\calr)}\right]^{-1}\,. 
\eeq 
For the particular form of Eq.~(\ref{eq:ricci}) we have 
\beq 
w_0\to\frac{1-4r_0}{3(1-\om)}\,. 
\eeq 
This model has one EOS parameter in addition to the matter density.  
We can therefore explore constraints 
either in the general kinematic plane $r_0$-$r_1$, or view them in the 
$\om$-$w_0$ plane.  Figure~\ref{fig:geom1} shows both.

Good complementarity, as well as concordance, exists among the probes 
in the $r_0$-$r_1$ plane.  One obtains an excellent fit with 
$(r_0,r_1)=(0.81,-0.72)$.  The value of $\calr$ today, $r_0$, approaches 
unity, the deSitter value.  Recall that $\calr=1/4$ corresponds to 
matter domination, and $\calr=1/2$ to the division between decelerating 
and accelerating expansion, so this kinematic approach clearly indicates 
the current acceleration of the universe. 

An interesting point to note is that $\lcdm$ is not a subset of 
this ansatz, i.e.\ the physics is distinct.  No values of $r_0$ and $r_1$ 
give a $\lcdm$ cosmology.  However, the Hubble diagram 
for the best fit agrees with that for $\lcdm$ to within 0.006 mag out to 
$z=2$ and 0.3\% in the reduced distance to CMB last scattering.   
This is especially interesting as this geometric dark energy model 
is almost purely kinematic.  The agreement 
appears in the $\om$-$w_0$ plane as contours tightly concentrated 
around $w_0=-1$, despite there being no actual scalar field or 
cosmological constant.  Again we note the excellent complementarity 
between the individual probes, even in this very different model. 

\subsection{Ricci dark energy ${\rm R_{high}}$} 

Rather than expanding the spacetime curvature around the present value 
we can also consider the deviation from a high redshift matter dominated 
era.  That is, we start with a standard early universe and ask how 
the data favors acceleration coming about.  In this 
second geometric dark energy model (call it ${\rm R_{high}}$ for high 
redshift or large values of scalar curvature), the value of $\calr$ 
evolves from $1/4$ at high redshift.  From the definition of $\calr$, 
it must behave asymptotically as 
\beq 
\calr=\frac{1}{4}\,\left[1-3\winf\frac{\delta H^2}{H^2}\right]\approx 
\frac{1}{4}\,[1+4\alpha\,a^{-3\winf}], 
\eeq 
where $\delta H^2=(H^2/H_0^2)-\om\,(1+z)^3$ is the deviation from matter 
dominated behavior, and $\winf$ is the associated, effective equation of 
state at high redshift, approximated as asymptotically constant.  

Next we extend this behavior to a form that takes the reduced scalar 
curvature to a constant in the far future (as it must if the 
EOS of the dominant component goes to an asymptotic value): 
\beq 
\calr=\frac{1}{4}+\frac{\alpha\,a^{-3\winf}}{1+\beta\,a^{-3\winf}}. 
\eeq 
So today $\calr=1/4+\alpha/(1+\beta)$ and in the future $\calr=1/4+\alpha/ 
\beta$.  By requiring the correct form for the high redshift Hubble 
expansion, one can relate the parameters $\alpha$ and $\beta$ by 
\beq 
\alpha=(3\beta\winf/4)[\ln\om/\ln(1+\beta)]\,,
\eeq 
and finally 
\beq 
H^2/H_0^2=\om\,a^{-3}\,(1+\beta a^{-3w_\infty})^{-\ln\om/\ln(1+\beta)}. 
\eeq 

The ${\rm R_{high}}$ geometric dark energy model has two parameters 
$\beta$ and $w_\infty$, 
in addition to the matter density $\om$.  This is the first such model 
we consider, and all remaining models also have two EOS parameters.  
Although current data cannot in general satisfactorily 
constrain two parameters, and so for all remaining models we do 
not show individual probe constraints, if the EOS phase space behavior 
of the model is sufficiently restrictive then reasonable joint 
constraints may result. 

\begin{figure}[!ht]
\begin{center}
\psfig{file=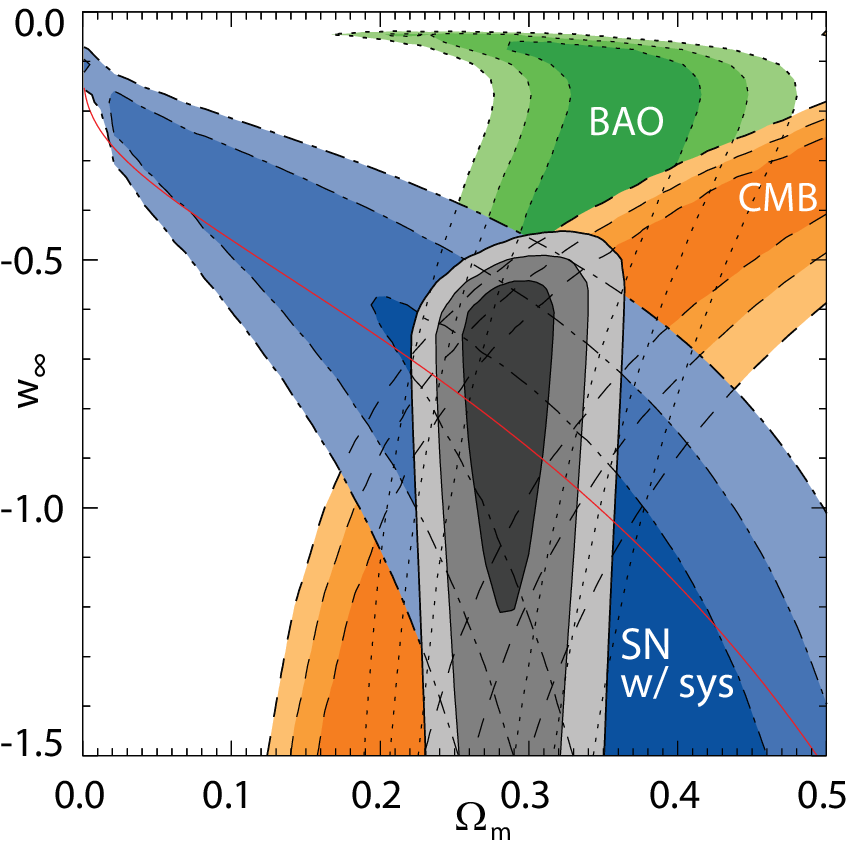, width=3.2in}
\psfig{file=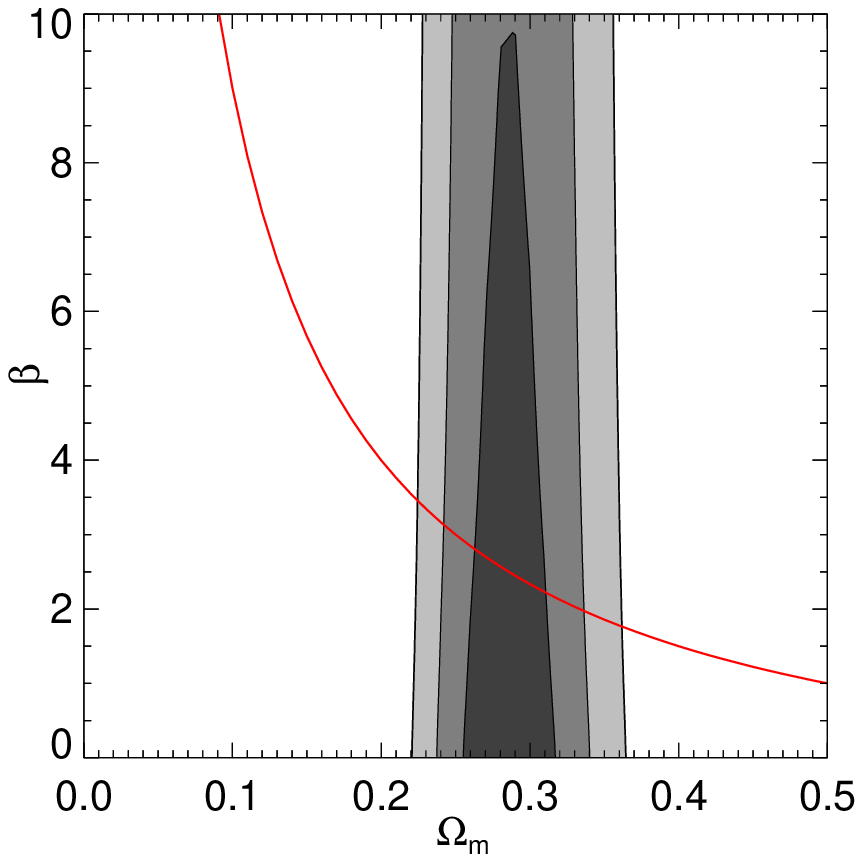, width=3.2in}
\caption{Geometric dark energy in the ${\rm R_{high}}$ model describes 
the acceleration 
directly through the reduced Ricci scalar curvature and deviations from 
early matter domination.  The left panel shows the $\om$-$w_\infty$ plane, 
indicating the nature of the deviation ($w_\infty=0$ corresponds to no 
transition away from matter domination), and the right panel shows the 
$\om$-$\beta$ plane, indicating the rapidity and fate of the deviation. 
The curve in the left panel corresponds to whether the fate of the universe 
is de Sitter; we also show the individual probe 
constraints, fixing $\beta$ to the de Sitter value (not minimizing over 
$\beta$ as for the joint contour), to show that SN closely 
map the fate of the universe.  In the right panel the curve is the cut 
through parameter space, fixing $\winf=-1$, corresponding to $\lcdm$. 
}
\label{fig:geom2}
\end{center}
\end{figure}

Figure~\ref{fig:geom2} shows the joint likelihoods in the $\om$-$w_\infty$ 
and $\om$-$\beta$ planes, with the third parameter minimized over.  We 
see that the data are consistent with the cosmological constant behavior 
$w_\infty=-1$ in the past (this is only a necessary, not sufficient 
condition for $\lcdm$), and indeed constrain the asymptotic high redshift 
behavior reasonably well, in particular to negative values of $w_\infty$.  
This indicates that the Ricci scalar curvature definitely prefers a 
nearly-standard early matter dominated era, i.e.\ the deviations faded 
away into the past.  This has important implications as well for 
scalar-tensor theories that would modify the early expansion history; 
in particular, the data indicate that deviations in $\calr$ must go 
approximately as $a^3$ (see \citet{lincahn}) not as $a$ as sometimes 
assumed. 

The parameter $\beta$ helps determine the rapidity of the Ricci scalar 
transition away from matter domination.  This varies between $\beta=0$, 
giving a slow transition but one reaching a deceleration parameter 
$q=-\infty$ in the asymptotic future, and $\beta\gg1$, giving a rapid 
deviation but with smaller magnitude.  A cosmological constant behavior 
has $\beta\approx 3$, as discussed below.

Within the three dimensional parameter space, two subspaces are of special 
interest.  One is where $w_\infty=-1$, a necessary condition for consistency 
with $\Lambda$, as mentioned.  The other corresponds to a deSitter 
asymptotic future, defined by the line 
\beq 
\beta_{\rm deS}=\om^{w_\infty}-1. \label{eq:geom2fate} 
\eeq 
Note that unlike the previous geometric model ${\rm R_{low}}$, the 
${\rm R_{high}}$ model does include $\Lambda$ as the limit when both 
these conditions are satisfied, $w_\infty=-1$ and $\beta=\om^{-1}-1$.  
This relation for the $\Lambda$ limit is shown as a curve in the 
$\om$-$\beta$ plane.  There is an overlap with the joint data 
likelihood, though one must be careful since the contours have been 
minimized over $\winf$. 

Interestingly, we can actually use the data to test consistency with a 
de Sitter asymptotic future.  This is shown by the curve in the 
$\om$-$\winf$ plane.  We see that SN are the probe most sensitive to testing 
the fate of the universe, with the SN contour oriented similarly to the 
curve given by Eq.~(\ref{eq:geom2fate}) that passes through the best fit.  
Thus the data are consistent with $w_\infty=-1$ and with a de Sitter 
fate separately, though some tension exists between satisfying them 
simultaneously.  Thus, this geometric dark energy may be distinct 
from the cosmological constant.

\section{PNGB Model} 

Returning to high energy physics models for dark energy, one of the key 
puzzles is how to prevent quantum corrections from adding a Planck energy 
scale cosmological 
constant or affecting the shape of the potential.  This is referred to 
as the issue of technical naturalness.  Pseudo-Nambu Goldstone boson (PNGB) 
models are technically natural, due to a shift symmetry, and so can 
be considered strongly physically motivated (perhaps even more so than 
$\Lambda$).  See \citet{friemanwaga} for an early cosmological analysis 
of PNGB as dark energy and more recent work by 
\citet{sorbo0612457,abrahamse0712.2879}.   

The potential for the PNGB model is 
\beq 
V(\phi)=V_\star\,[1+\cos(\phi/f)], 
\eeq 
with $V_\star$ setting the magnitude, $f$ the symmetry energy scale or 
steepness 
of the potential, and $\phi_i$ is the initial value of the field when it 
thaws from the high redshift, high Hubble drag, frozen state.  These three 
parameters determine, and can be thought of as roughly analogous to, the 
dark energy density, the time variation of the equation of state, and the 
value of the equation of state.  The dynamics of this class of models 
is sometimes approximated by the simple form 
\beq 
w(a)=-1+(1+w_0)a^F, 
\eeq 
with $F$ roughly inversely related to the symmetry energy scale $f$, but 
we employ the exact numerical solutions of the field evolution equation. 

PNGB models are an 
example of thawing dark energy, where the field departs recently from 
its high redshift cosmological constant behavior, evolving toward a less 
negative equation of state.  Since the EOS only deviates recently from 
$w=-1$, the precision in measuring $w_0$ is more important than the 
precision in measuring an averaged or pivot EOS value.  SN data provide 
the tightest constraint on $w_0$.  In the future the field oscillates around 
its minimum with zero potential and ceases to accelerate the expansion, 
acting instead like nonrelativistic matter.  

\begin{figure}[!ht]
\begin{center}
\psfig{file=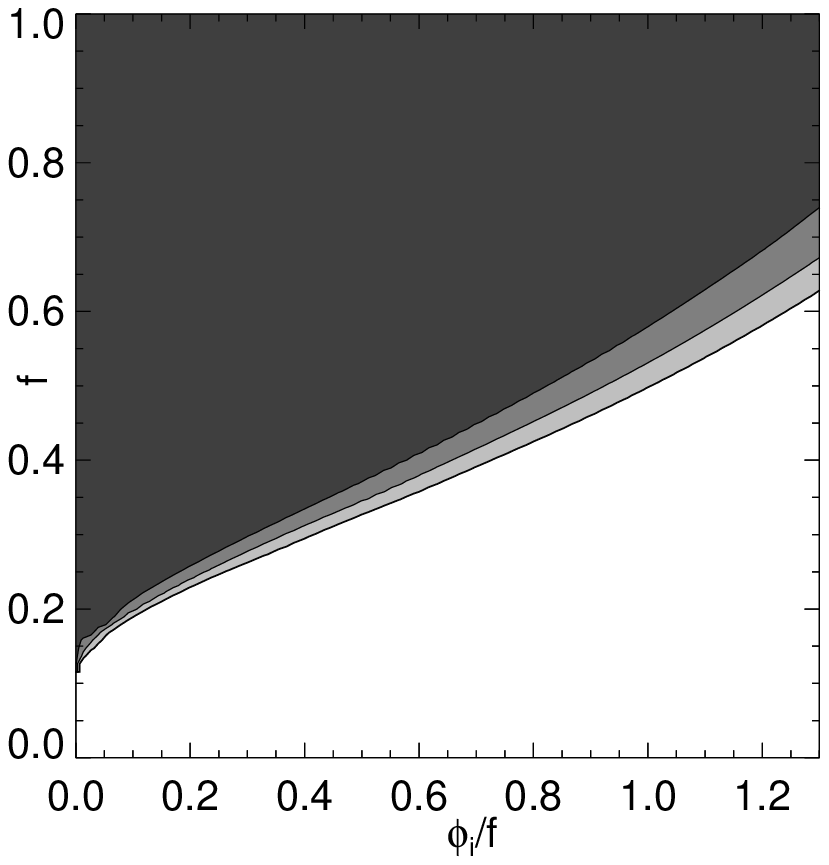, width=3.2in} 
\psfig{file=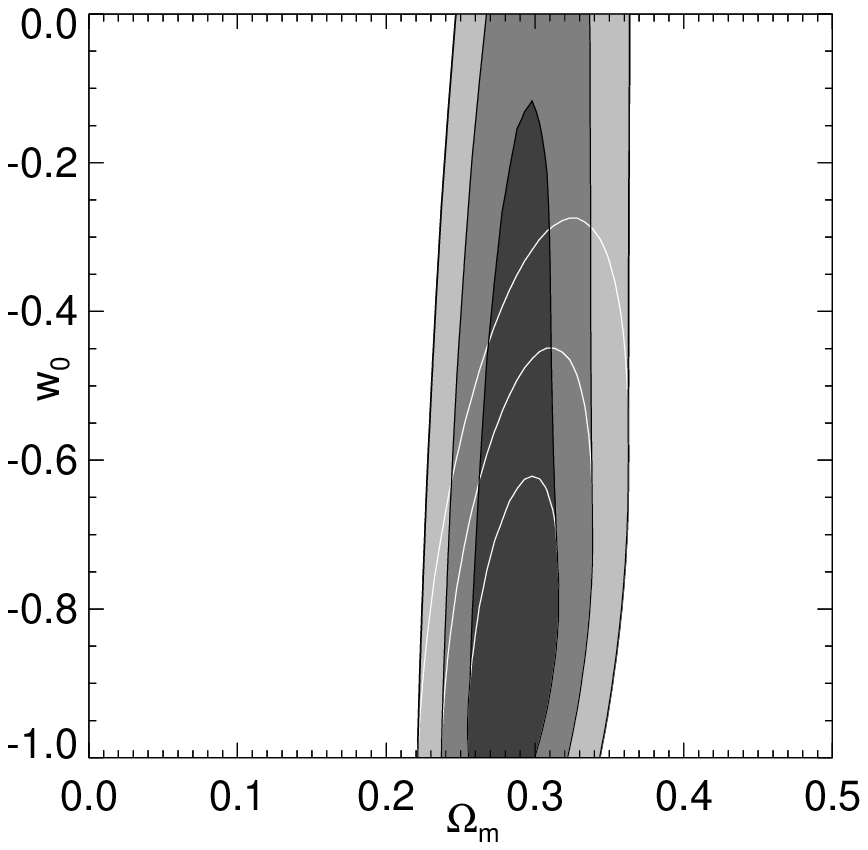, width=3.2in} 
\caption{{\bf Left panel:} PNGB model dynamics involves a competition 
between the steepness of the potential, given by the symmetry energy 
scale $f$, and the initial field position $\phi_i/f$.  
If the potential is very steep, $f\ll1$, the field will roll 
so rapidly to the potential minimum that the dark energy density never 
becomes significant, unless $\phi_i/f$ is fine tuned very near zero.  For 
natural energy scale values near the Planck scale, $f\approx1$, a wide 
variety of $\phi_i/f$ are viable. 
{\bf Right panel:} The field spends a long period frozen, acting as a 
cosmological constant before thawing and evolving to a present EOS 
$w_0$.  For steep potentials with $f\ll1$, the thawing can be rapid 
and result in evolution to $w_0$ far from $-1$, yet still be consistent 
with data.  The solid confidence 
level contours in the $w_0$-$\om$ plane show PNGB results for 
energy scales $f\ge0.1$, while the white outline contours consider 
only PNGB models with more natural energy scales $f\ge0.5$; the 
latter favors models closer to the cosmological constant behavior. 
} 
\label{fig:pngb}
\end{center} 
\end{figure}

Figure~\ref{fig:pngb} illustrates the constraints in both the 
particle physics and cosmological parameters.  The symmetry energy scale 
could provide a key clue for revealing the fundamental physics 
behind dark energy, and it is interesting to note that these astrophysical 
observations essentially probe the Planck scale.  For values of $f$ 
below unity (the reduced Planck scale), the potential is steeper, 
causing greater evolution away 
from the cosmological constant state.  However, the field may be frozen 
until recently and then quickly proceed down the steep slope, allowing 
values of $w_0$ far from $-1$ but looking in an average or constant $w$ 
sense like $\langle w\rangle\approx-1$.  Small values of $\phi_i/f$ have 
the field set initially ever more finely near the top of the 
potential; starting from such a flat region the field rolls very little 
and $w$ stays near $-1$ even today.  In the limit $\phi_i/f=0$ the field 
stays at the maximum, looking exactly like a cosmological constant.  
The two effects of the steepness and initial position mean that the 
cosmological parameter likelihood can accommodate both $w_0\approx-1$ and 
$w_0$ approaching 0 as consistent with current data.  However, to agree 
with data {\it and\/} $1+w_0\sim1$ requires $f\ll1$ and fine tuning -- 
e.g.\ for $f=0.1$ one must balance the field to within one part in a 
thousand of the top. 
Thus in the left panel there exists an invisibly narrow tail extending 
along the y-axis to $f=0$.  In the right panel, we show how taking 
more natural values $f\gtrsim0.5$ removes the more extreme values of 
$w_0$ caused by the unnatural fine tuning.

\section{Algebraic Thawing Model} 

While PNGB models involve a pseudoscalar thawing field, we can also consider 
scalar fields with thawing behavior.  Any such fields that are neither 
fine tuned nor have overly steep potentials must initially depart from the 
cosmological constant behavior along a specific track in the EOS 
phase space, characterized by a form of slow roll behavior in the matter 
dominated 
era.  (See \citet{caldlin,paths,schsen,cahndl}.)  Here we adopt the algebraic 
thawing model of \citet{grg}, specifically designed to incorporate this 
physical behavior: 
\beqa 
1+w&=&(1+w_0)\,a^p\,\left(\frac{1+b}{1+ba^{-3}}\right)^{1-p/3} \\ 
H^2/H_0^2&=&\om\,a^{-3}+(1-\om)\,\exp 
\left[\frac{3(1+w_0)}{\alpha p}\left\{1-(1-\alpha+\alpha a^3)^{p/3}\right\}\right], 
\eeqa 
where $\alpha=1/(1+b)$ and $b=0.3$ is a fixed constant not a parameter.  
The two parameters are $w_0$ and $p$ and this form fulfills the physical 
dynamics condition not only to leading but also next-to-leading order 
\citep{cahndl}. 

\begin{figure}[!ht]
\begin{center}
\psfig{file=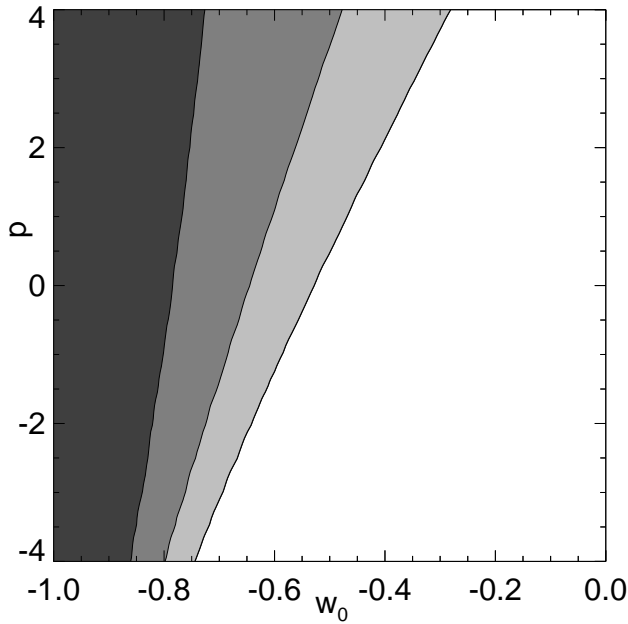, width=3.4in} 
\caption{Algebraic thawing model incorporates the expected physical 
behavior of a thawing scalar field rolling slowly from a matter dominated 
era.  Such a model is a fairly generic parametrization for this class 
of physics when $p\in[0,3]$, and has a strong goodness of fit. 
} 
\label{fig:algthaw}
\end{center} 
\end{figure}

The physical behavior of a minimally coupled scalar field evolving from a 
matter dominated era would tend to have $p\in[0,3]$.  Since we want to 
test whether the data points to such a thawing model, we consider values 
of $p$ outside this range.  Results are shown in Figure~\ref{fig:algthaw}. 

For $p<0$, the field has already evolved to its least negative value of $w$ 
and returned toward the cosmological constant.  The more negative $p$ is, 
the less negative (closer to 0) the extreme value of $w$ is, so these 
models can be more tightly constrained as $p$ gets more strongly negative.  
As $p$ gets more 
positive, the field takes longer to thaw, increasing its similarity to 
the cosmological constant until recently, when it rapidly evolves to $w_0$. 
Such models will be very difficult to distinguish from $\Lambda$.  
If we restrict consideration to the physically expected range 
$p\in[0,3]$, this implies $w_0<-0.57$ at 95\% confidence in these thawing 
models, so considerable dynamics remains allowed under current data. 
This estimation is consistent with the two specific thawing models 
already treated, the doomsday and PNGB cases. 

The goodness of fit to the data is the best of all models considered 
here, even taking into account the number of fit parameters.  This 
may indicate that we should be sure to include a cosmological probe 
sensitive to $w_0$ (not necessarily the pivot EOS $w_p$) and to 
recent time variation $w_a$, such as SN, in our quest to understand 
the nature of dark energy.

\section{Early Dark Energy} 

The other major class of dark energy behavior is that of freezing models, 
which start out dynamical and approach the cosmological constant in 
their evolution.  The tracking subclass is interesting from the point of 
view again of fundamental physics motivation: they can ameliorate the 
fine tuning problem for the amplitude of the dark energy density by having 
an attractor behavior in their dynamics, drawing from a large basin of 
attraction in initial conditions \citep{zws}.  Such models generically 
can have 
nontrivial amounts of dark energy at high redshift; particularly interesting 
are scaling models, or tracers, where the dark energy has a fixed fraction 
of the energy density of the dominant component.  These can be motivated 
by dilatation symmetry in particle physics and string theory \citep{wett88}. 

As a specific model of such early dark energy we adopt that of 
\citet{DoranRobbers0601544}, with 
\beq 
\omde(a)=\frac{1-\om-\ome\,(1-a^{-3w_0})}{1-\om+\om a^{3w_0}} 
+\ome\,(1-a^{-3w_0}) \label{eq:heid}
\eeq
for the dark energy density as a function of scale factor $a=1/(1+z)$.  
Here $\omde=1-\om$ is the present dark energy density, $\ome$ is the 
asymptotic early dark energy density, and $w_0$ is the present dark 
energy EOS.  In addition to the matter density the two parameters are 
$\ome$ and $w_0$.  

The Hubble parameter is given by $H^2/H_0^2=\om\,a^{-3}/[1-\omde(a)]$. 
The standard formula for the EOS, 
$w=-1/(3[1-\omde(a)])\,d\ln\omde(a)/d\ln a$, does not particularly 
simplify in this model.  Note that 
the dark energy density does not act to accelerate expansion at early 
times, and in fact $w\to0$.  However, although the energy density 
scales like matter at high redshift, it does not appreciably clump and 
so slows growth of matter density perturbations.  We will see this 
effect is crucial in constraining early dark energy. 

\begin{figure}[!ht]
\begin{center}
\psfig{file=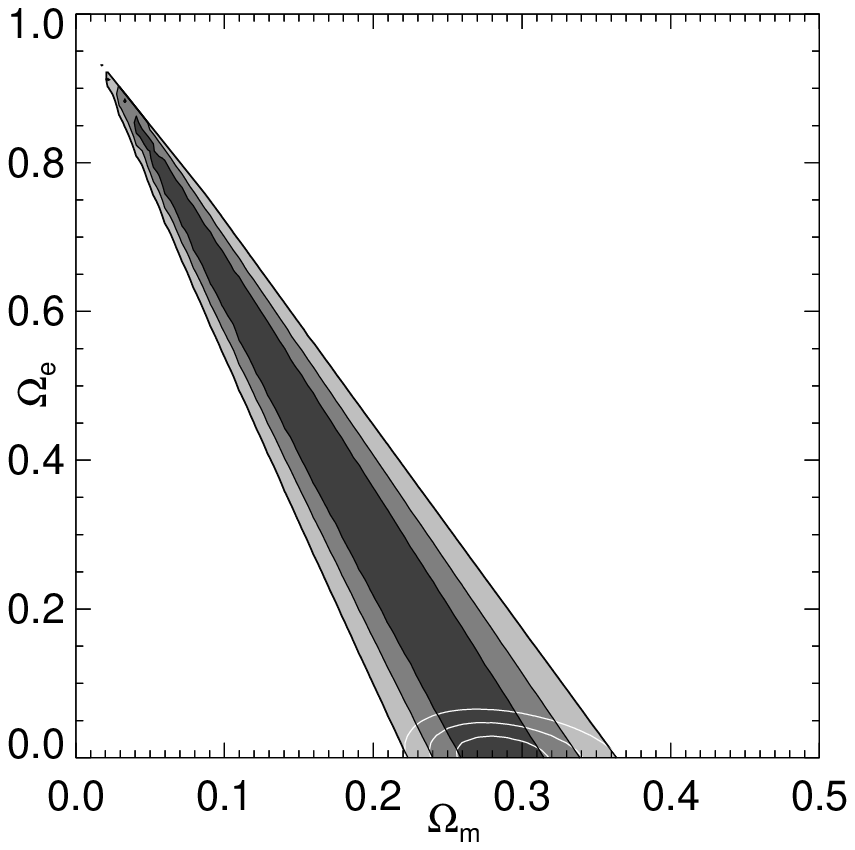, width=3.2in}
\psfig{file=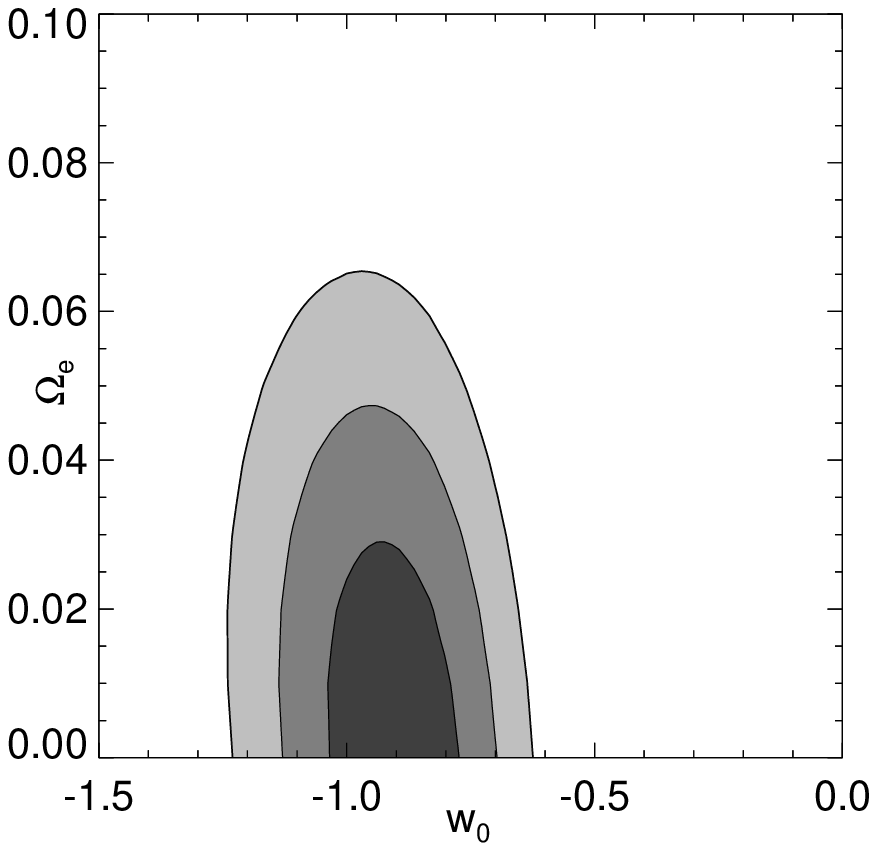, width=3.2in}
\caption{Early dark energy represents an example of a freezing model 
with interesting particle physics motivations.  The left panel shows 
the constraints on $\ome$ and $\om$ from purely geometric data, as used 
throughout this article.  The degeneracy evident in the contours leaves 
the acoustic scales unchanged, but hides the shift in the sound horizon 
caused by early dark energy, leading to possible misinterpretation of 
the correct cosmological model.  The degeneracy can be broken by adding 
growth information, here a 10\% prior on total linear growth (or 
$\sigma_8$), as shown by the white outline contours.  
This tightly restricts the early dark energy density to contribute no 
more than a few percent.  The right panel shows the $\ome$-$w_0$ 
constraints including the growth prior. 
}
\label{fig:early}
\end{center}
\end{figure}

Figure~\ref{fig:early} shows the constraints in the $\om$-$\ome$ and 
$\ome$-$w_0$ planes.  
Considerable early dark energy density appears to be 
allowed, but this is only because we used purely geometric information, 
i.e.\ distances and the acoustic peak scale.  The high redshift Hubble 
parameter for a scaling solution is multiplied by a factor $1/\sqrt{1-\ome}$ 
relative to the case without early dark energy (see \citet{doranst}). 
This means that the sound horizon is shifted according to 
$s\sim\sqrt{1-\ome}$, but a geometric degeneracy exists whereby the 
acoustic peak angular scale can be preserved by changing the value of 
the matter density $\om$ (see \citet{linrobb} for 
a detailed treatment).  This degeneracy is clear in the left panel. 

However, as mentioned, the growth of perturbations is strongly affected 
by the unclustered early dark energy.  This suppresses growth at early 
times, leading to a lower mass amplitude $\sigma_8$ today.  To explore 
the influence of growth constraints, we investigate adding a growth prior
of 10\% to the data, i.e.\ we require the total linear growth (or 
$\sigma_8$) to lie within 10\% of the concordance model.  
The innermost, white contour of the left panel of Fig.~\ref{fig:early} 
shows the constraint with the growth prior.  In the right panel we zoom 
in, and show $\ome$ vs.\ $w_0$, seeing that the degeneracy is effectively 
broken.  The amount of early dark energy is limited to $\ome<0.038$ at 
95\% cl. 
Similar conclusions were found in a detailed treatment by 
\citet{doranwett0609814}.  

We find a convenient fitting formula is that for an early dark energy 
model the total linear growth to the present is suppressed by 
\beq 
\frac{\Delta g_0}{g_0}\approx \left(\frac{\ome}{0.01}\right)\times 5.1\%, 
\eeq 
relative to a model with $\ome=0$ but all other parameters fixed.   
Thus appreciable amounts of early dark energy have significant effects on 
matter perturbations, and we might expect nonlinear growth to be even more 
sensitive (e.g.\ see \citet{bartel}).

\section{Growing Neutrino Model \label{sec:grownu}} 

While freezing or scaling models such as the early dark energy model 
just considered are interesting from the physics perspective, they 
generically have difficulty in evolving naturally to sufficiently 
negative EOS by the present.  The growing neutrino model of 
\citet{Amendola0706.3064,Wetterich0706.4427} solves this by coupling the 
scalar field to massive neutrinos, forcing the scalar field to a 
near cosmological constant behavior when the neutrinos go nonrelativistic. 
This is an intriguing model that solves the coincidence problem through 
cosmological selection (the time when neutrinos become nonrelativistic) 
rather than tuning the Lagrangian. 

The combined dark sector (cosmon scalar field plus mass-running neutrinos) 
energy density is 
\beqa 
\omds(a)&=&\frac{\omds a^3+2\omv(a^{3/2}-a^3)}{1-\omds(1-a^3)+2\omv(a^{3/2} 
-a^3)}, \qquad a>a_t \\ 
\omds(a)&=&\ome, \qquad\qquad\qquad\qquad\qquad a<a_t \,, 
\eeqa 
where $\omds=1-\om$ is the present dark sector energy density.  The Hubble 
parameter can be found by $H^2/H_0^2=\om a^{-3}/[1-\omds(a)]$ as usual.  
The two free dark parameters are the neutrino mass or density 
$\omv=m_\nu(z=0)/(30.8h^2\,{\rm eV})$ and the early dark energy density 
$\ome$.  The transition scale factor $a_t$ is determined by intersection 
of the two behaviors given for $\omds(a)$.  

The equation of state is 
\beq 
w=-1+\frac{\omv a^{-3/2}}{\omds+2\omv (a^{-3/2}-1)}, \qquad a>a_t 
\eeq 
with $w=0$ before the transition, i.e.\ a return to the standard early 
dark energy model.  One can therefore translate $\omv$ or $m_\nu(z=0)$ 
into $w_0=-1+\omv/\omds=-1+\omv/(1-\om)$.  

\begin{figure}[!ht]
\begin{center}
\psfig{file=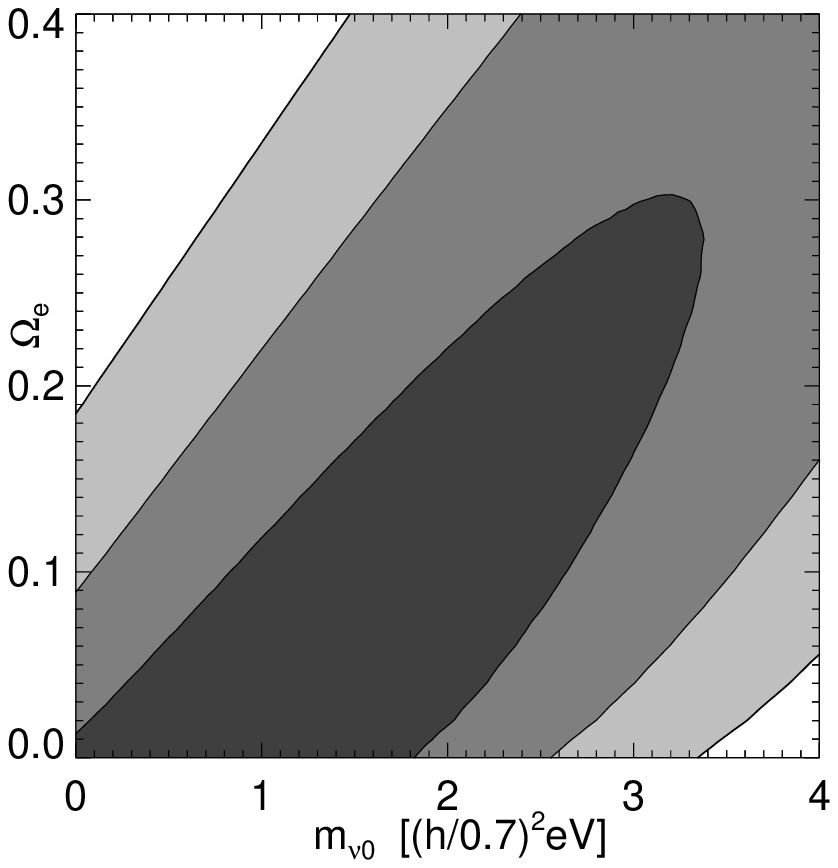, width=3.2in}
\psfig{file=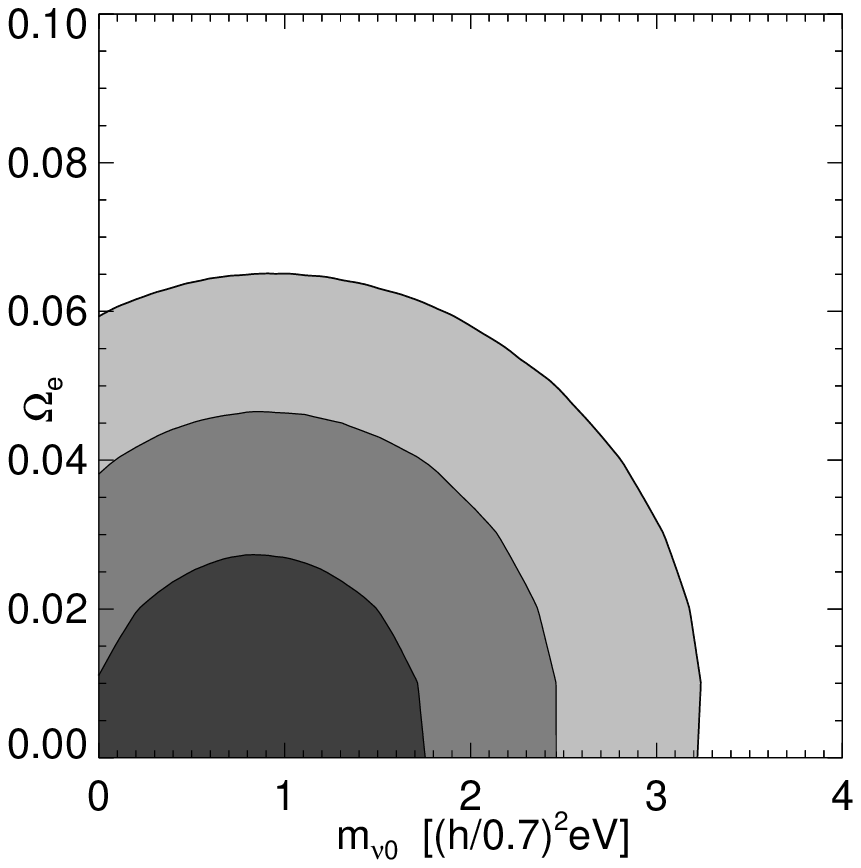, width=3.2in}
\caption{Growing neutrino model, coupling a dark energy scalar field 
to massive neutrinos, can solve the coincidence problem.  The left panel 
shows the constraints from purely geometric data, while the right panel 
(note the different vertical scale) 
adds a 10\% prior on total linear growth (also see Fig.~\ref{fig:early}). 
The neutrino mass today becomes tightly constrained to an interesting 
range, and comparison with laboratory limits could lead to evidence of 
varying neutrino mass. 
}
\label{fig:neut}
\end{center}
\end{figure}

Figure~\ref{fig:neut} shows the constraints in the $m_\nu(z=0)$-$\ome$ 
plane.  As in the previous early dark energy model, the geometric 
degeneracy is clear.  Again, when we add growth information in the form 
of a 10\% prior on the total linear growth (or the mass variance 
$\sigma_8$), the constraints tighten considerably, as shown in the 
right panel.  The 95\% confidence 
level limit on the neutrino mass from this current cosmological data 
is then $2.1\,(h/0.7)^2$ eV ($1.2$ if only statistical uncertainties are 
taken into account).  These limits are comparable to astrophysical 
constraints from similar types of data applied to standard, constant 
mass neutrinos \citep{goobarnu,tegmarknu}.  Note that because the 
neutrino mass grows due to the coupling, the value today can actually 
be larger than that at, say, $z\approx3$ where Lyman alpha forest 
constraints apply \citep{seljak}.

\section{Conclusion \label{sec:concl}} 

We have considered a wide variety of dark energy physics quite different 
from the cosmological constant.  These include a diversity of physical 
origins for the acceleration of the expansion: from dynamical scalar 
fields to dark energy that will eventually cause deceleration and collapse, 
to gravitational modifications arising from extra dimensions or from 
quantum phase transitions, to geometric or kinematic parametrization 
of the acceleration, to dark energy that may have influenced the early 
universe and that may have its magnitude set by the neutrino mass. 
The comparison to $\lcdm$ and constant $w$ cases covers 5 one-parameter 
and 5 two-parameter dark energy equation of state models. 
(\citet{lh05} detail how even next generation data will not generically 
be able to tightly constrain more than two such parameters.) 

Two key results to emphasize are that current data 1) are consistent with 
$\Lambda$, and 2) are also consistent with a diversity of other models and 
theories, even when we restrict consideration to those with at least 
modest physical motivation or justification.  As explicitly shown by 
the mirage model, any inclination toward declaring $\lam$ the answer 
based on consideration of a constant $w$ has an overly restricted view.  
The need for next generation observations with far greater accuracy, and 
the development of precision growth probes, such as weak gravitational 
lensing, is clear.  All major classes of physics to explain the 
nature of dark energy are still in play.  

However there are already quite hopeful signs of imminent progress in 
understanding the nature of dark energy.  For example, for the braneworld 
model tight control of systematics would decrease the goodness of fit 
to $\Delta\chi^2=+15$, even allowing for spatial curvature, diminishing 
its likelihood by a factor 2000 na{\"\i}vely, effectively 
ruling out the model.  For the doomsday model, improving errors by 30\% 
extends our ``safety margin'' against cosmic collapse by 10 billion years -- 
a nonnegligible amount!  Every improvement in uncertainties pushes the 
limits on the neutrino mass within the growing neutrino model closer 
toward other astrophysical constraints -- plus this model essentially 
guarantees a deviation from $w=-1$ of $0.1\,(m_{\nu 0}/{\rm eV})$, 
excitingly tractable.  Terrestrial neutrino oscillation bounds already 
provide within this model that $1+w>0.005$. 

As points of interest, we note that the model  
with noticeably positive $\Delta\chi^2$ relative to $\Lambda$, 
and hence disfavored, is completely distinct from the 
cosmological constant, i.e.\ the braneworld model has no limit within 
its parameter spaces equivalent to $\Lambda$.  This does not say that 
no such model could fit the data -- the $R_{\rm low}$ model is also 
distinct from $\Lambda$ but fits as well as many models.  Certainly 
many successful models under current data do look in some averaged sense 
like a vacuum energy but this does not necessarily point to static 
dark energy.  Two serious 
motivations to continue looking for deviations are that physicists have 
failed for 90 years to explain the magnitude required for a cosmological 
constant, and that the previous known occurrence of cosmic acceleration -- 
inflation -- evidently involved a dynamical field not a cosmological 
constant. 

To guide further exploration of the possible physics, we highlight 
those models which do better than $\Lambda$: 
the geometric dark energy and algebraic thawing approaches.  One of the sole 
models where adding a degree of freedom is justified (albeit modestly) 
by the resulting reduction in $\chi^2$ is the ${\rm R_{high}}$ model 
directly studying deviations of 
the spacetime curvature from the matter dominated behavior.  This has one 
more parameter than the constant $w$ EOS approach, but improves in $\chi^2$ 
by 1.  In addition, it has a built-in test for the asymptotic de Sitter 
fate of the future expansion.  We recommend that this model be considered 
a model of interest for future fits.  The other model improving by at 
least one unit of $\chi^2$ is the algebraic thawing model, performing 
better than the other thawing models, with a general parametrization 
explicitly incorporating the physical conditions imposed by matter 
domination on the scalar field dynamics. 

The diversity of models also illustrates some properties of the 
cosmological probes beyond the familiar territory of vanilla $\lcdm$. 
For example, for the algebraic thawing and other such evolutionary models, 
the premium is on precision of $w_0$ and $w_a$ much more than the averaged 
or pivot EOS value $w_p$.  Not all models possess the wonderful 
three-fold complementarity of the probes seen in the constant $w$ case; 
for many of the examples BAO and CMB carry much the same information as 
each other.  However, we clearly see that 
for every model SN play a valuable role, complementary to CMB/BAO, and 
often carries the most important physical information: such as on the 
doomsday time or the de Sitter fate of the universe or the Planck scale 
nature of the PNGB symmetry breaking. 

The diversity of physical motivations and interpretations of acceptable 
models highlights the issue of assumptions, or priors, on how the dark 
energy should behave.  For example, in the $R_{\rm low}$ model should 
priors be flat in $r_0$, $r_1$ or in $\om$, $w_0$; in the PNGB model 
should they be flat in $f$, $\phi_i/f$ or in $\om$, $w_0$, etc.? 
Lacking clear physical understanding of the appropriate priors restricts the 
physical meaning of any Bayesian evidence one might calculate 
to employ model selection; the $\chi^2$ goodness of fit used here does 
not run into these complications that can obscure physical interpretation. 

We can use our diversity of models for an important consistency test of 
our understanding of the {\it data\/}.  If there would be systematic trends 
in the data which do not directly project into the $\lcdm$ parameter space 
(i.e.\ look like a shift in those parameters), then one might expect that 
one of the dozen models considered might exhibit a significantly better fit.  
The fact that we do not observe this can be viewed as evidence that the 
data considered here is not flawed by significant hidden systematic 
uncertainties.  The data utilize the Union08 compilation of uniformly 
analyzed and crosscalibrated Type Ia supernovae data, constituting the 
world's published set, with systematics treated and characterized 
through blinded controls.  The data are publicly available at 
http://supernova.lbl.gov/Union, and will be supplemented as further 
SN data sets become published; the site contains high resolution figures 
for this paper as well.  

However, to distinguish deeply among the 
possible physics behind dark energy requires major advances in several 
cosmological probes, enabling strong sensitivity to the time variation 
of the equation of state.  This is especially true for those models 
that are now or were in the past close to the cosmological constant 
behavior.  We are getting our first glimpses looking beyond $\Lambda$, 
but await keen improvements in vision before we can say we 
understand the new physics governing our universe.

\acknowledgments 

We thank Andy Albrecht, Robert Caldwell, Roland de Putter, Steven 
Weinberg, and Christof Wetterich for helpful discussions.  
This work has been supported in part by the Director, Office of Science, 
Office of High Energy Physics, of the U.S.\ Department of Energy under 
Contract No.\ DE-AC02-05CH11231. 
M.K. acknowledges support from the Deutsche Forschungsgemeinschaft (DFG).

\end{document}